\documentclass[preprint2]{aastex}
\usepackage{color}

\shorttitle{Solar Small Scale Microwave Bursts}
\shortauthors{Baolin Tan}

\begin{document}

\title{Small Scale Microwave Bursts in Long-duration Solar Flares}

\author{Baolin Tan$^1$} \affil{$^1$ Key Laboratory of Solar Activity,
National Astronomical Observatories of Chinese Academy of Sciences,
Beijing 100012, China} \email{bltan@nao.cas.cn}

\begin{abstract}

Solar small scale microwave bursts (SMBs), including microwave
dot, spike, and narrow band type III bursts, are characterized
with very short timescales, narrow frequency bandwidth, and very
high brightness temperatures. Based on observations of the Chinese
Solar Broadband Radio Spectrometer at Huairou with superhigh
cadence and frequency resolution, this work presents an intensive
investigation of SMBs in several flares occurred in active region
NOAA 10720 during 2005 Jan 14-21. Especially long-duration flares,
SMBs occurred not only in early rising and impulsive phase, but
also in the flare decay phase, and even in time of after the flare
ending. These SMBs are strong bursts with inferred brightness
temperature at least $8.18\times10^{11}$ - 1.92$\times10^{13}$ K,
very short lifetime of 5-18 ms, relative frequency bandwidths of
0.7-3.5\%, and superhigh frequency drifting rates. Together with
their obviously different polarizations from the background
emission (the quiet Sun, and the underlying flaring broadband
continuum), such SMBs should be individual independent strong
coherent bursts which is related to some non-thermal energy
releasing and production of energetic particles in small scale
source region. These facts show the existence of small scale
strong non-thermal energy releasing activities after the flare
maxima, which is meaningful to the prediction of space weather.
Physical analysis indicates that plasma mechanism may be the most
favorable candidate for the formation of SMBs. From plasma
mechanism, the velocities and kinetic energy of fast electrons can
be deduced, and the region of electron acceleration can also be
tracked.

\end{abstract}

\keywords{Sun: activity --- Sun: flares --- Sun: particle emission
--- Sun: radio radiation}

\section{Introduction}

In solar broadband spectral observations at microwave frequency
range, especially at decimeter and centimeter wavelength, it has
been known for a long time that there are many isolated
short-timescale strong bursts, such as microwave spike bursts
(Benz, 1986; Huang \& Nakajima, 2005; Rozhansky et al, 2008;
Debrowski et al, 2011), dot bursts (Karishan et al. 2003), and
narrow bandwidth type III bursts (Huang \& Tan, 2012) occurred
frequently associated with solar flares. The main features of
these bursts include very short timescales (lifetime $<0.1$s),
very narrow frequency bandwidth (relative bandwidth $<5\%$), very
high brightness temperatures ($T_{b}>10^{12}K$), and always
occurred in great clusters. Practically, there are always no
obviously differences among these three types of bursts, we may
call them by a joint name as solar small scale microwave bursts
(SMBs), which possibly associates with similar strong non-thermal
energy releasing processes. SMBs are always superposed on the
underlying broadband flaring continuum emission (Malville, Aller,
\& Jensen, 1967, etc) and frequently occurred in the rising and
maximum phases of solar impulsive flares (Beaz, 1986, Csillaghy \&
Benz, 1993, etc). However, recent observations indicate that they
also occurred in flare decay phase and even after the flare ending
(Benz et al, 2002; Huang \& Tan, 2012).

The obvious differences of SMBs from the underlying broadband
flaring continuum emissions indicate that they may have entirely
different generation mechanisms. Actually, SMBs are possibly
related to some strong non-thermal processes with very short
timescale and very small size of flaring magnetic energy release
region. Possibly, these processes may be elementary in solar
flares, which should be fragmented (Benz, 1985; Messmer \& Benz,
2000). However, so far, there is no perfect model to explain the
formation of flaring fragmentation and the generation of SMBs. As
for the source locations of SMBs, there are two different
viewpoints of theories. The electron cyclotron maser emission
mechanism (ECME) predicted that the SMB emission would be expected
to originate from some loss-cone instabilities in some regions
close to the footpoints of magnetic loop with strong magnetic
field: $\omega_{ce}>>\omega_{pe}$ (Melrose \& Dulk 1982). Here,
$\omega_{ce}$ is electron gyro-frequency and $\omega_{pe}$ is
electron plasma frequency. In other theories, SMBs are proposed to
produce from acceleration processes and therefore would be
expected to originate in the location of acceleration sites. Tan
\& Tan (2012) found that each pulse of a microwave quasi-periodic
pulsation at frequency of 2.60-3.80 GHz is composed of a group of
SMBs, and the concomitant zebra pattern indicates that the
corresponding magnetic field strength is only 147-210 Gs, which is
too weak to meet the ECME conditions. Therefore, they proposed
that SMBs are triggered by plasma emission, another coherent
emission mechanism, which is triggered by Langmuir turbulence
produced from non-thermal energetic electrons (Zheleznyakov \&
Zlotnik 1975; Chernov et al. 2003). In this case, the formation of
SMBs is closely related to electron accelerations and the
evolution of plasma instability. Benz et al. (2002) find that
decimetric spikes are single source occurred in the flare decay
phase and located about 20-400$''$ away from the flare site in
hard X-ray and soft X-ray bursts. This fact indicates that the
decimetric spikes are questionable to related with the main flare
electron acceleration, but possibly with some coronal post-flare
acceleration processes. Therefore, the much more detailed
investigation of SMBs will give insight into the elementary
processes taking place in flaring regions, and reveal the
intrinsic properties of accelerations of non-thermal particles.

In order to recognize SMBs clearly, telescopes need enough high
spectral resolutions and high cadence. Gudel-Benz law indicates
that the lifetime of SMBs is largely a function of emission
frequency in frequency range 237-2695 MHz: $\tau\propto
f^{-1.29\pm0.08}$ (Gudel \& Benz, 1990). This law predicts that
the lifetime around 1.20 GHz is about 12 ms. Csillaghy \& Benz
(1993) found that bandwidth of individual SMB is also a function
of emission frequency: $\triangle f\sim0.66f^{0.42}$, which
implies the frequency bandwidth about 13 MHz at frequency around
1.20 GHz. The Chinese Solar Broadband Radio Spectrometer in
Huairou (SBRS/Huairou) at 1.10-1.34 GHz has a cadence of 1.25 ms
and spectral resolution of 4 MHz. It is sufficient to recognize
SMBs occurred in the corresponding frequency range.

This work presents a comprehensive investigation of SMBs by using
the broadband spectral observations obtained by SBRS/Huairou in
several solar flares occurred in active region NOAA 10720 during
2005 January 15-20. A large number of SMBs are registered in these
flares. Section 2 introduces the observation data and the main
properties and evolutionary characteristics of SMBs in these
flares. Section 3 presents a physical discussion on the above
results, and finally, some conclusions are summarized in Section
4.

\section{Observations and Analysis}

\subsection{The Flare Events}

The solar active region NOAA 10720 is a flare-productive and most
impressive one on the solar visible disk in January of 2005. It
appeared on the solar disk as a simple beta type magnetic region
on Jan. 10, grew rapidly and fully developed on Jan 15, ended as a
complex sunspot region on Jan. 23. There are 5 X-class flares, 17
M-class flares, and more than 60 C-class flares produced in this
active region and recorded by GOES satellites during its 14 days
track across the solar disk (from 2005 January 10 to January 23).
Cheng et al. (2011) investigated the properties of these flares
and their relationships to coronal mass ejections (CME). Zhao \&
Wang (2006) investigated the non-potentiality and free energy
transportation in this active region.

\begin{deluxetable}{cccccccccccccc} 
\tablecolumns{9} \tabletypesize{\scriptsize} \tablewidth{0pc}
\tablecaption{List of solar flares observed by SBRS/Huairou in
NOAA 10720 during 2005 Jan 14-21\label{tbl-1}} \tablehead{
 \colhead{Date}& \colhead{Flare}&\colhead{start(UT)}&\colhead{peak(UT)}&\colhead{end(UT)}&\colhead{Duration(min)}&\colhead{GOES Class}&\colhead{$N_{smb}$}& \colhead{NorP} &\\}
 \startdata
    Jan15 &  E1   &   00:22    &   00:43    &  01:02    &   40           & X1.2        &     90    &  Yes   \\
          &  E2   &   03:16    &   03:40    &  03:57    &   41           & C4.2        &      0    &  No    \\
          &  E3   &   04:09    &   04:16    &  04:22    &   13           & M1.3        &      0    &  No    \\
          &  E4   &   04:26    &   04:31    &  04:36    &   10           & M8.4        &      0    &  Yes   \\
          &  E5   &   05:54    &   06:37    &  07:17    &   83           & M8.6        &   6000    &  Yes   \\\hline
    Jan17 &  E6   &   03:10    &   03:21    &  03:32    &   22           & M2.6        &      0    &  Yes   \\\hline
    Jan18 &  E7   &   00:37    &   00:44    &  00:51    &   14           & C6.0        &      0    &  No    \\
          &  E8   &   02:06    &   02:12    &  02:17    &   11           & C3.2        &      0    &  No    \\\hline
    Jan19 &  E9   &   05:10    &   05:26    &  05:35    &   25           & C7.2        &      0    &  No    \\
          &  E10  &   06:58    &   07:31    &  07:55    &   57           & M6.7        &     45    &  No    \\
          &  E11  &   08:03    &   08:22    &  08:40    &   37           & X1.3        &     92    &  No    \\\hline
    Jan20 &  E12  &   03:21    &   03:30    &  03:36    &   15           & C4.8        &     17    &  Yes   \\
          &  E13  &   06:36    &   07:01    &  07:26    &   50           & X7.1        &  12600    &  Yes   \\
\enddata
\tablecomments{$N_{smb}$ is the total number of SMBs occurred in
each flare at frequency of 1.10-1.34 GHz. The last column
indicates whether or not complete observations by Nobeyama Radio
Polarimeters (NorP).}
\end{deluxetable}

SBRS/Huairou observed 13 events perfectly at frequency of
1.10-1.34 GHz and 2.60-3.80 GHz around the occurrence of the
active region NOAA 10720 during 2005 January 14-21, including 3
X-class flares, 5 M-class flares, and 5 C-class flares. Table 1
listed the main parameters of these 13 flare events. There are 3
events belonging to long-duration flares (duration $\geq$ 50 min)
including 2 M-class flares (E5, E10) and 1 X7.1 flare (E13); 7
events belong to short-duration (duration $\leq$ 30 min) including
3 M-class flares (E3, E4, E6) and 4 C-class flares (E7, E8, E9,
E12). Cheng et al (2011) reported that E1 (X1.2 flare), E4 (M8.4
flare) and E6 (M2.6 flare) are confined flares without CMEs, while
the long-duration events E5, E10, and E13 are impulsive flare with
strong CMEs.

\subsection{Observation Data}

Here, the observation data is obtained by SBRS/Huairou, which is a
group of advanced high-performance solar radio spectrometers with
high cadence, broad frequency bandwidth, and high frequency
resolution (Fu et al. 1995, 2004, Yan et al. 2002). SBRS/Huairou
includes 3 parts: 1.10 - 2.06 GHz (the antenna diameter is 7.0 m,
cadence is 5 ms, frequency resolution is 4 MHz. When it works at
1.10-1.34 GHz, the cadence is 1.25 ms), 2.60 - 3.80 GHz (the
antenna diameter is 3.2 m, cadence is 8 ms, frequency resolution
is 10 MHz), and 5.20 - 7.60 GHz (share the same antenna of the
second part, cadence is 5 ms, frequency resolution is 20 MHz).
Antennae point to the center of solar disk automatically
controlled by computers. The spectrometer can receive the total
flux of solar radio emission with dual circular polarization
(left- and right-handed circular polarization, LCP and RCP), and
the dynamic range is 10 dB above quiet solar background emission.
The observation sensitivity is: $\delta F\leq 2\%S_{\bigodot}$,
here $S_{\bigodot}$ is the standard flux of the quiet Sun. From
the Solar Geophysical Data (SGD) we can obtain $S_{\bigodot}$ at
frequencies of 610 MHz, 1415 MHz, 2695 MHz, 2800 MHz, and 4995
MHz, and make the calibration of the observational data followed
the method reported by Tanaka et al. (1973). As for the strong
burst, the receiver may work beyond its linear range and a
nonlinear calibration method is replaced (Yan et al. 2002).
Similar to other congeneric instruments, such as Phoenix (100 -
4000 MHz, Benz et al. 1991), Ond\'rejov (800 - 4500 MHz, Jiricka
et al. 1993) and Brazilian Broadband Spectrometer (BBS, 200 - 2500
MHz, Sawant et al. 2001), SBRS/Huairou has no spatial resolution.
However, as the Sun is a strong radio emission source, a great
deal of works (e.g. Dulk 1985, etc.) indicate that the microwave
bursts received by spectrometers are always coming from the solar
active region when the antenna points to the Sun. A specialized
software based on IDL program has also been developed for
analyzing the spectrograms, which can display the spectral fine
structures clearly. In order to corroborate the results obtained
by SBRS/Huairou, we also scrutinize the observation data obtained
by Nobeyama Radio Polarimeters (NorP). We found that NorP also got
the completed observations during the several interesting flare
events, such as E1, E4, E5, E6, E12, and E13. This work focused on
studying these flare events.

In order to present a receivable statistical analysis of SMBs, it
is necessary to confirm practical definitions of parameters for
describing SMBs. Here we follow the definitions mainly from
Tarnstrom \& Philip (1972) to obtain the parameters.

(1) Burst strength ($F$): defined as the maximum emission
intensity subtracting the background emission intensity before and
after the corresponding SMB. A flux enhancement $F$ which exceeds
obviously to the quiet Sun emission may be regarded as a microwave
burst. An universal accepted criterion is 5$\sigma$, here $\sigma$
is the standard deviation of the background emission, when a flux
enhancement exceeds $F\geq5\sigma$ respect to the background
emission, we may say it is an isolated burst event. For example,
From the SGD record during 2005 Jan 15-20, we may get the standard
flux value of the quiet Sun $S_{\bigodot}\simeq 30-32$ sfu at
frequency of 1.10 - 1.34 GHz. The corresponding sensitivity is
about $\delta F\sim 0.60-0.64$ sfu.

Because the signal of SMB is much shorter timescale and much
narrower bandwidth than that of the underlying flare continuum
emission, it is easily affected by other factors, such as the
instrument perturbations. Here we adopt the similar criteria of
high energy physical experiments of 5 sigma to show a real
observation result (3-5 sigma to show an evidence, and $<3$ sigma
only reflects a hint or clue). Naturally, this criteria will leave
out some SMBs with relatively weak intensities. However, such high
criteria can help us to avoid the influence of other factors and
increase the confidence level of our analysis.

(2) Lifetime ($\tau$): can be measured from time interval of the
SMB temporal profile with half maximum of emission intensity.

(3) Frequency bandwidth ($\triangle f$): can be measured from
frequency width of the SMB spectral profile with half emission
intensity maximum.

In the above three parameters, the key step is to determine the
time or frequency at the maximum value and the half maximum value.
Here we use an IDL GAUSSFIT function fitting method to compute the
non-linear least-squares fit to a function as:

$F(t)=a_{0}+a_{1}e^{-b_{t}(t-t_{0})^{2}}$.

Then, the burst intensity is $F=a_{1}$, the averaged emission
intensity of the background is $a_{0}$, the time of maximum
intensity is at $t_{0}$, and SMB lifetime is
$\tau=2\sqrt{\frac{ln2}{b_{t}}}\simeq\frac{1.665}{\sqrt{b_{t}}}$.
The left panel of Fig.1 is an example of this method applying to
the temporal profile of an SMB. Here, the burst strength $F\simeq
35$ sfu, lifetime $\tau\simeq13.1$ ms.

As the averaged emission intensity of the background is increasing
with frequency, we may use another function to fit the spectral
profile of SMB:

$F(f)=a_{0}+a_{1}e^{-b_{f}(f-f_{0})^{2}}+a_{2}f$.

Here, $f$ is the emission frequency, $a_{0}+a_{2}f$ presents the
averaged emission intensity of the background at different
frequencies. The frequency bandwidth of SMB can be obtained
$\triangle
f=2\sqrt{\frac{ln2}{b_{f}}}\simeq\frac{1.665}{\sqrt{b_{f}}}$. The
right panel of Fig.1 is an example applying to the spectral
profile of an SMB. Here, the frequency bandwidth of SMB is
$\triangle f\approx14.9$ MHz.

\begin{figure*}[ht]   
   \includegraphics[width=7.6 cm]{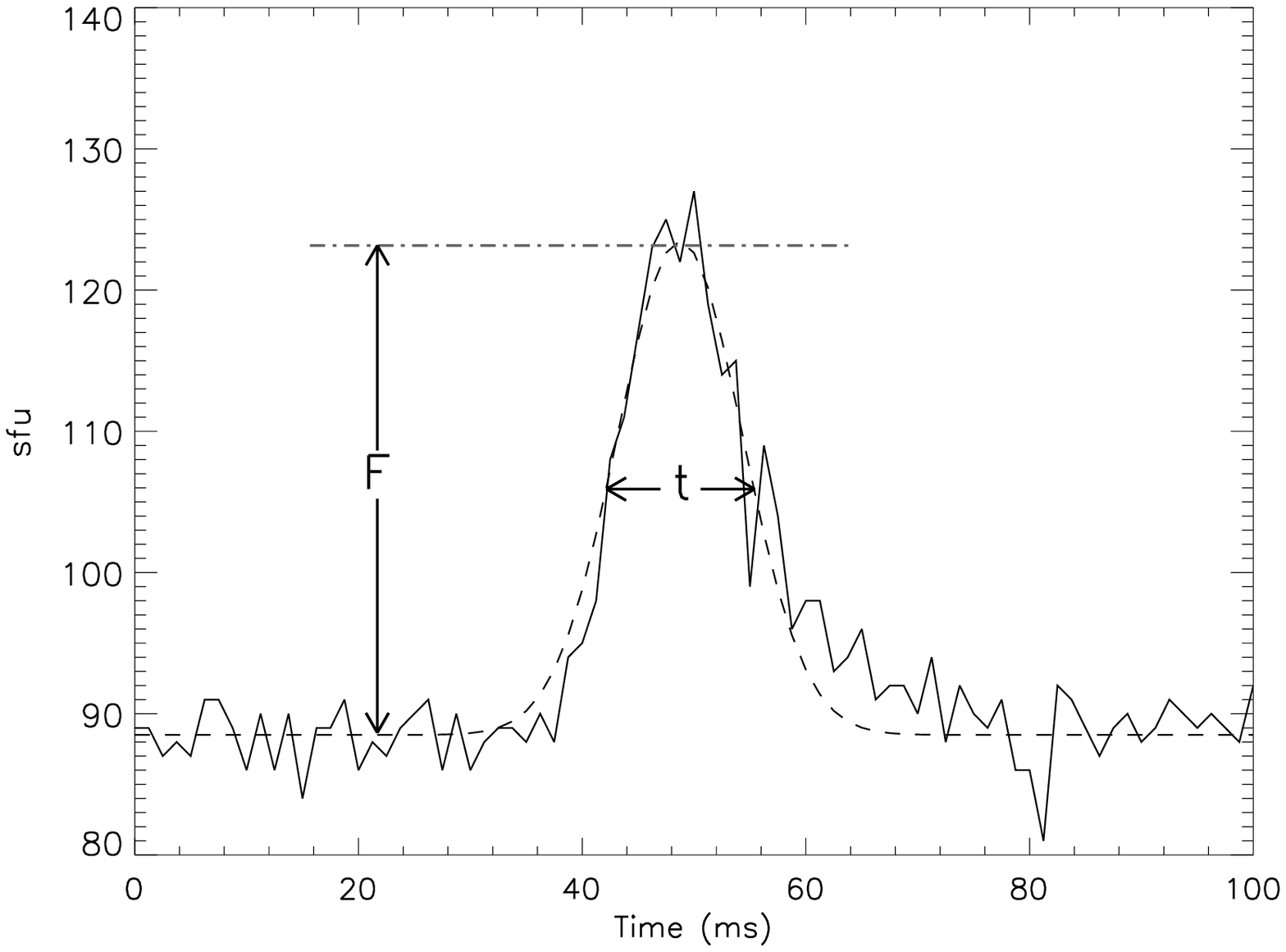}
   \includegraphics[width=7.6 cm]{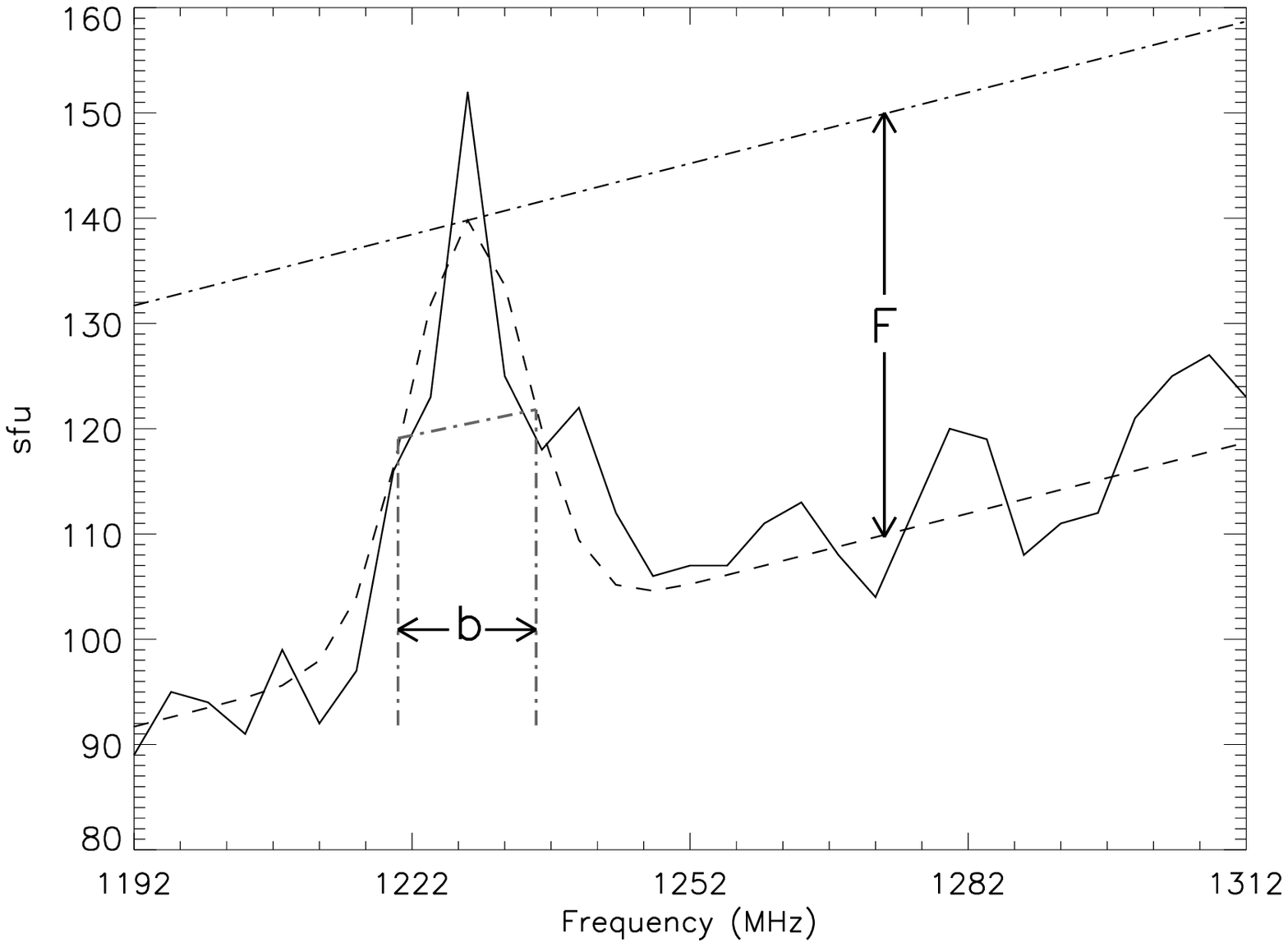}
\caption{Extracting burst strength (F), lifetime (t), and
frequency bandwidth (b) of SMBs by Gaussian function fitting
method. The solid curve is the original observation profile, the
dashed curve is the resulting profile of Gaussian function
fitting.} \label{fig:source}
\end{figure*}

Sometimes, SMBs become so crowded that we can't distinguish them
with each other. In order to make the result believable, here we
only analyze the isolated SMBs, while neglect the crowded SMBs
which can not be distinguished from each other clearly. The
isolated SMB is defined as: the quench time between two SMBs is
longer than its lifetime, and the frequency gap between two SMBs
is wider than its frequency bandwidth.

(4) Frequency drifting rate ($R_{drift}$): defined as the slope of
SMB on the time-frequency plane.

(5) Polarization degree, $r=\frac{F_{L}-F_{R}}{F_{L}+F_{R}}\times
100\%$. Here, $F_{L}$ and $F_{R}$ are burst strength of SMB at LCP
and RCP which subtract the background emissions, respectively.
$r>0$ indicates the left polarization, and $r<0$ indicates the
right polarization.

(6) SMB occurrence rate ($N_{smb}$), because almost all SMB in
this work are occurred only at frequency 1.10-1.34 GHz, therefore
we define $N_{smb}$ as the number of SMBs per second at frequency
of 1.10-1.34 GHz.

Table 1 lists the solar flares observed by SBRS/Huairou in NOAA
10720 during 2005 Jan 14-21. The following sections present the
comprehensive analyzing results based on the above definitions.

\subsection{Analysis Results}

From the observation of SBRS/Huairou at frequency 1.10-1.34 GHz
and 2.60-3.80 GHz, we found that:

(1) There are 6 flares occurring isolated SMBs, and all isolated
SMBs only appeared at frequency of 1.10-1.34 GHz. The observations
at frequency of 2.60-3.80 GHz obtained by SBRS/Huairou were also
inspected carefully, and found that there were no SMBs occurred in
the above flares, except some quasi-periodic pulsations (in E4,
E5, E9, and E11) and zebra pattern structures (E1 and E5 on Jan.
15). This fact imply that all SMBs occurred only in the frequency
band below 2.60 GHz in the active region NOAA 10720, which is
different from other cases, such as the flare occurred in active
region NOAA 10930 on 2006-12-13 (Wang, et al. 2008).

(2) The 8th column of Table 1 listed the number of isolated SMBs
occurring in each flare at frequency of 1.10-1.34 GHz. Here, we
find that different flare has different manifestations of SMBs:
(a) the most abundant SMBs are occurred in two long-duration
flares: E5, an M8.6 flare occurred on 2005-01-15 with duration of
83 min and about 6000 SMBs at frequency 1.10-1.34 GHz; and E13, an
X7.1 flare occurred on 2005-01-20 with duration of 50 min and
about 12600 SMBs; (b) some sporadic SMBs occurred in other 4
flares, including E1, E10, E11, and E12; (c) the short-duration
flares tend to be lack of SMBs, such as E2, E3, E4, E6, E7, E8,
and E9.

The following paragraphs present the detailed properties of SMBs
appeared in these flares.

\subsubsection{Abundant SMBs in a long-duration M8.6 flare on 2005-01-15}

The GOES soft X-ray emission indicates that the M8.6 flare starts
at 05:54 UT, reaches to maximum at 06:37 UT, and ends at 07:17 UT
on 2005-01-15 with duration of 83 min. It is a long-duration flare
accompanying with a powerful CME with speed up to more than 2000
km/s and abundant SMBs. SMBs appear not only in the early rising
and impulsive peak phases of the flare, but also occur in the
flare decay phase. Fig.2 presents two segments of SMBs which
occurred in the rising and decay phase of the long-duration M8.6
flare (E5) on 2005-01-15, respectively. The left panels show a
segment spectrograms of SMBs during 06:02:05.0 - o6:02:06.0 UT in
the flare rising phase with LCP (left-upper) and RCP
(left-bottom). The bright small patches represent SMBs which show
very strong left polarization, and distribute randomly. The
frequency bandwidth of individual SMB is in the range of 8-32 MHz,
and the lifetime is about 5-21 ms. As a comparison, the quench
time between each adjacent two SMBs is in the range of 50 - 100 ms
which is much longer than SMB lifetime. The frequency gap between
each adjacent two SMBs is in the range of 24-64 MHz, which is
wider than the frequency bandwidth of each individual SMB. The
right panels of Fig.2 presents another segment of SMBs during
06:38:22.0 - 06:38:23.0 UT in the flare decay phase. It shows that
almost all SMBs are appeared in RCP with strong polarization. Most
of the lifetimes are shorter and the frequency bandwidths are
narrower than that occurred in the flare rising phase. In both of
the above two cases, the frequency drifting rates of some SMBs can
be measured, and the values of $R_{drift}$ are in range of 1500 -
6000 MHz/s. At the same time, most of SMBs behavior as
perpendicular to the time axis that we cannot detect their
$R_{drift}$. Possibly this is because of their high frequency
drifting rates that we can not detect them for restrictions of
limited spectral and temporal resolutions. For example, when the
bandwidth is 12 MHz, the detectable maximum $R_{drift}$ is 9600
MHz/s for the cadence 1.25 ms, etc.

\begin{figure*}[ht]   
   \includegraphics[width=7.6 cm, height=8.8 cm]{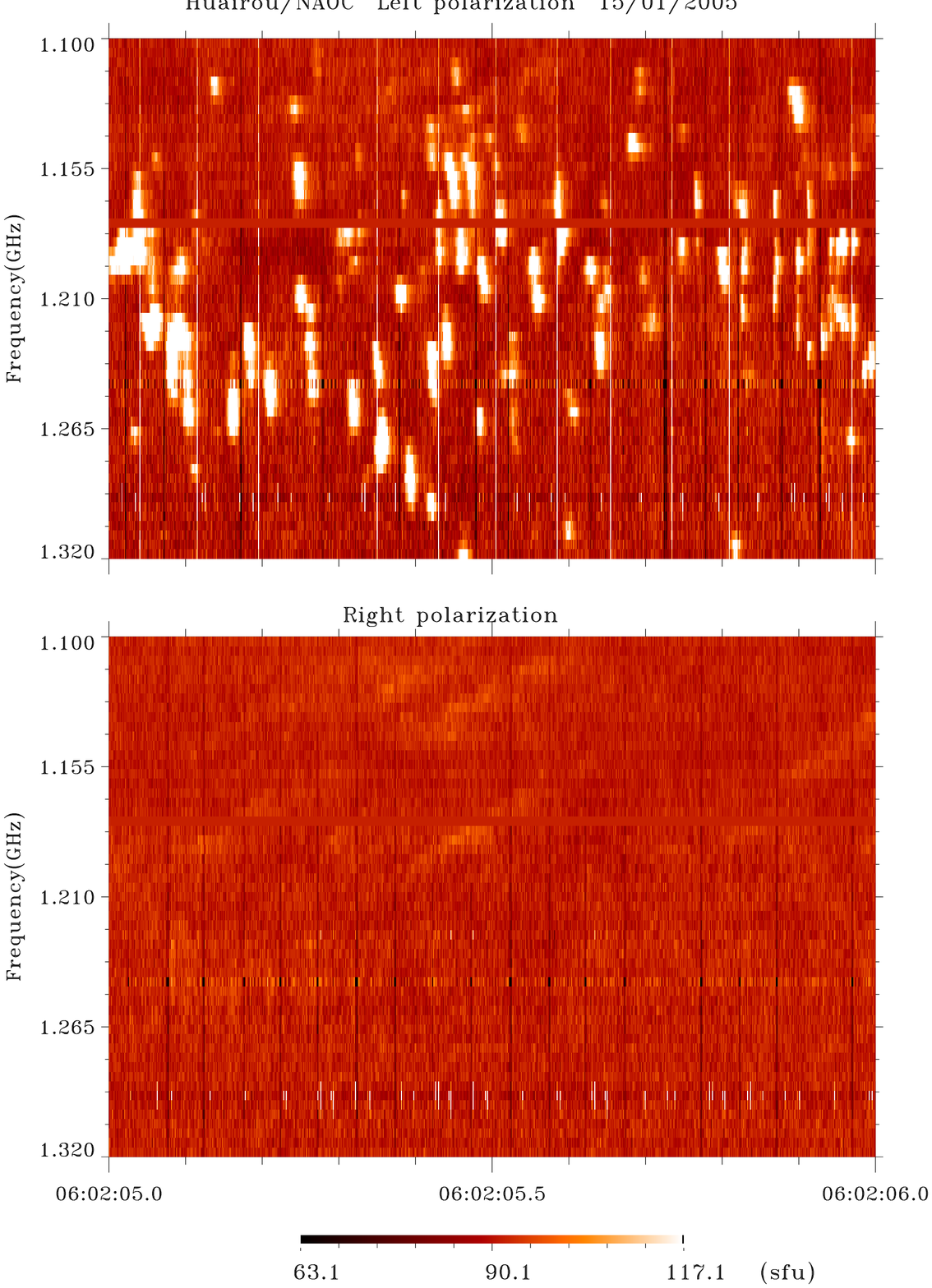}
   \includegraphics[width=7.6 cm, height=8.8 cm]{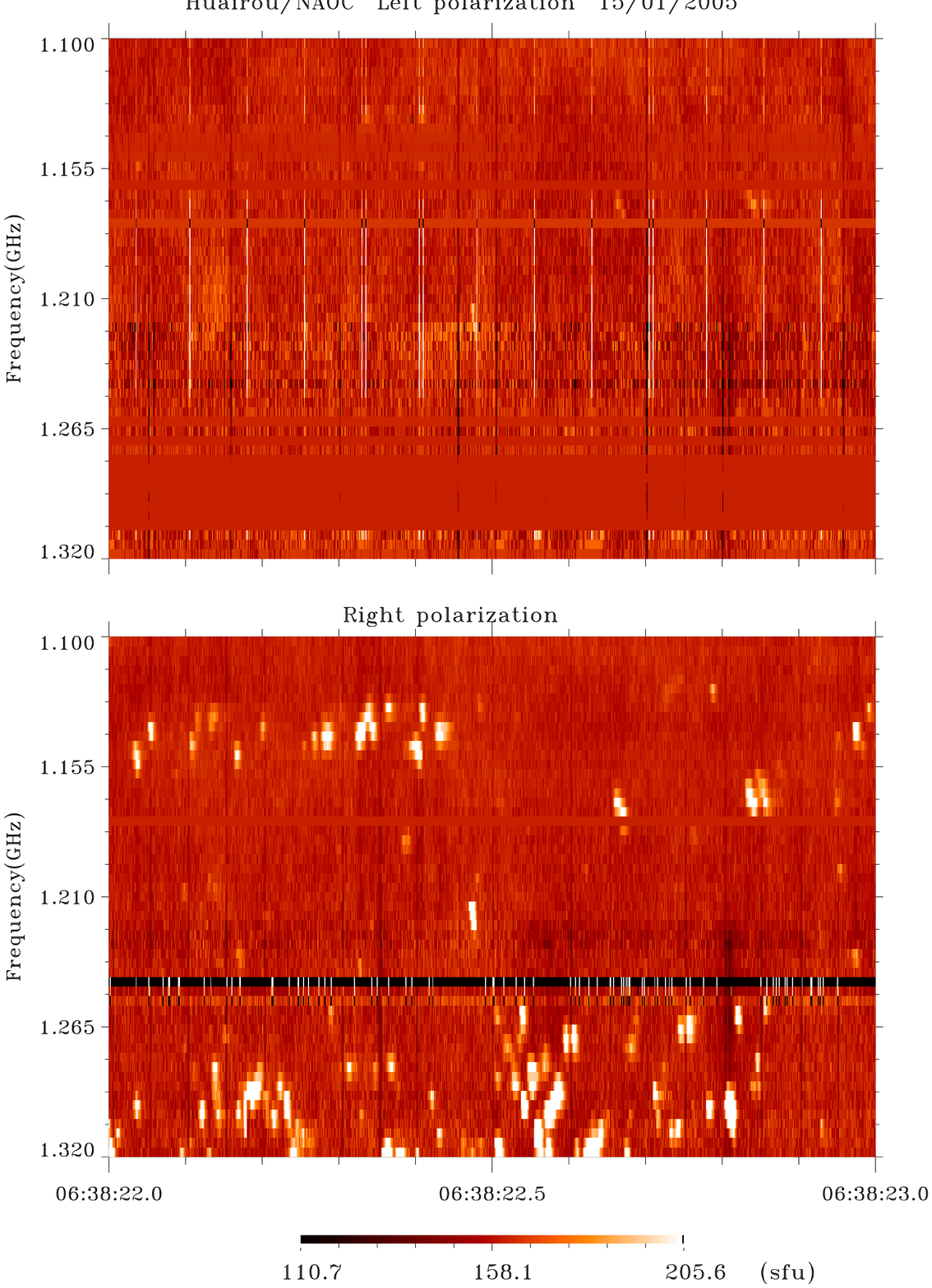}
\caption{The left and right panels present spectrograms of SMBs at
LCP and RCP observed by SBRS/Huairou in the rising and decay
phases of an M8.6 flare (E5) on 2005-01-15.} \label{fig:source}
\end{figure*}

Fig.3 presents the temporal profiles of emission intensities of
two segments of SMBs with LCP and RCP, respectively. As a
comparison, the emission intensity of the quiet Sun recorded
during 05:45:00.0-05:45:00.6 UT (before the onset of the flare) is
also over-plotted on the same figure. It is obvious that the
microwave emission can be decomposed into three components. The
first component is the quiet Sun emission ($F_{quiet}$) which can
be represented by the record before the solar flare (e.g. during
05:45 - 05:50 UT); the second one is the underlying broadband
flaring continuum emission ($F_{flare}$) which can be defined as
the emission during the gap between two SMBs; and the last one is
the emission of SMBs ($F_{SMB}$) which exceeds $F_{flare}$
significantly. Around the M8.6 flare, $F_{quiet}\sim 29-30$ sfu at
LCP and RCP without significant polarization. The left panel of
Fig.3 shows that $F_{flare}\sim 107.4$ sfu (the standard deviation
$\sigma=5.09$ sfu) at LCP and 87.7 sfu ($\sigma=2.74$ sfu) at RCP,
and $F_{SMB}$ at LCP is in the range of 43-85 sfu which exceeds
5$\sigma$ with respect to $F_{flare}$ and far surpass the
instrument sensitivity $\delta F$. The right panel of Fig.3
indicates that $F_{flare}\sim 153.7$ sfu ($\sigma=4.85$ sfu) at
LCP and 157.8 sfu ($\sigma=3.26$ sfu) at RCP. $F_{SMB}\sim 42-145$
sfu at RCP which also exceeds 5$\sigma$ with respect to
$F_{flare}$ and far surpass the instrument sensitivity $\delta F$.
We may suppose that each SMB is an independent burst which is
different from the underlying broadband flaring continuum and the
quiet Sun emissions.

\begin{figure*}[ht]  
   \includegraphics[width=7.6 cm]{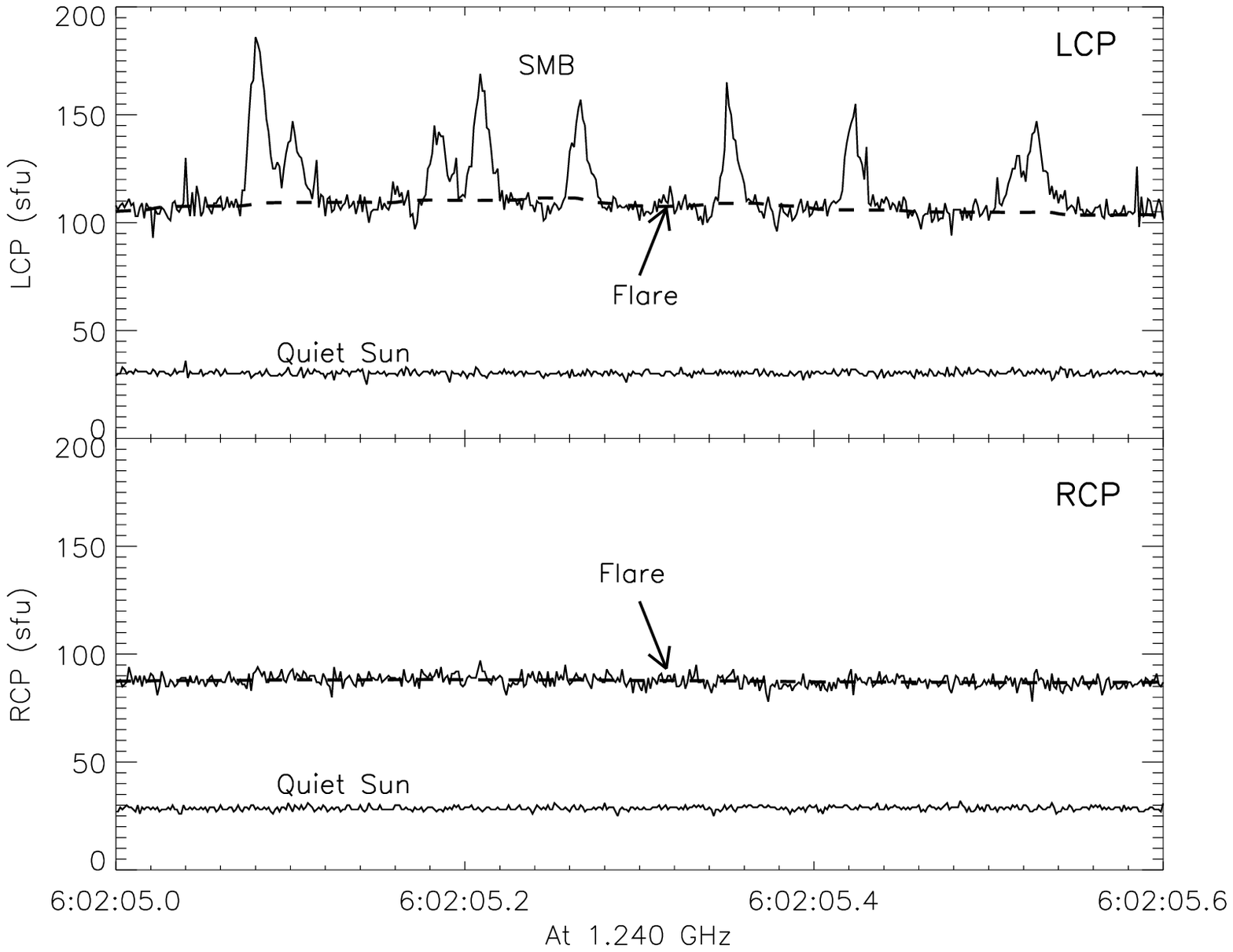}
   \includegraphics[width=7.6 cm]{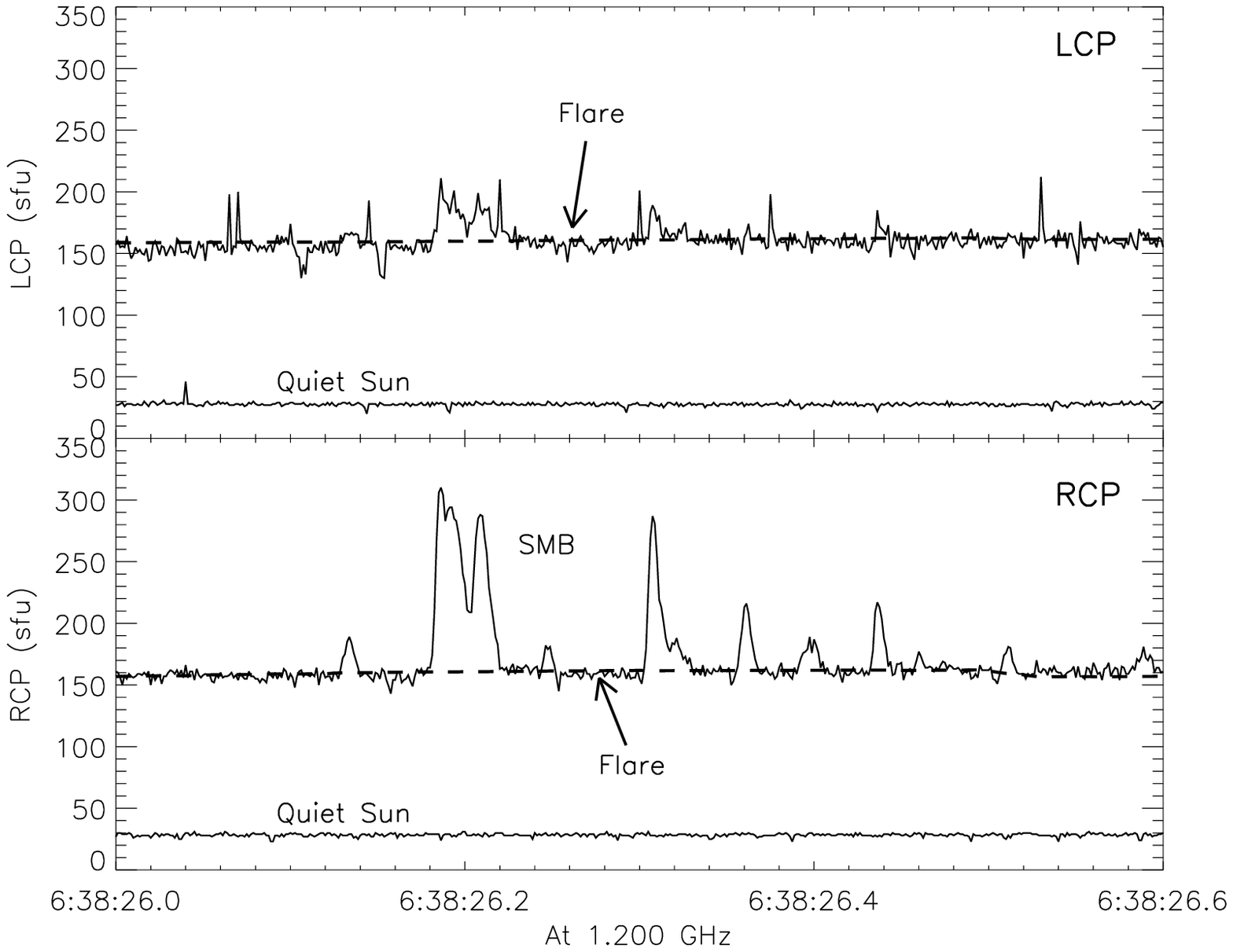}
\caption{Comparisons among emission components of the quiet Sun,
flare and SMBs in two segments of SMBs occurred in the flare
rising (left) and decay (right) phases of an M8.6 flare (E5) on
2005-01-15. Here, the quiet Sun emission is recorded during
05:45:00.0-05:45:00.6 UT, before the onset of the flare.}
\label{fig:source}
\end{figure*}

The spectrogram in Fig. 2 implies that the distribution of SMBs is
randomly, either in the flare rising phase or in the decay phase.

Actually, from the onset of the long-duration M8.6 flare to its
decay phase, there are numerous SMBs occurred. The left panel of
Fig.4 presents the distribution of the SMB occurrence rate in the
flare and a comparison with the GOES soft X-ray emission (GOES
SXR), the microwave emission intensity at LCP and RCP. The plus
signs (+) indicate SMB occurrence rate. The solid curves are the
radio emission intensity at 1.20 GHz. The positive and negative
values indicate the presence at LCP and RCP, respectively. The
comparison among GOES SXR, the microwave emission intensity at LCP
and RCP indicate that there are time differences between their
maxima. The maximum of GOES SXR occurred at about 06:37 UT, of LCP
at about 06:25 UT, and of RCP at about 06:16 UT. A rough
estimation indicates that there are about 6000 SMBs associated
with the flare. Among them, about 94\% LCP SMBs occurred before
the LCP maximum, and about 91\% RCP SMBs occurred after the LCP
maximum.

\begin{figure*}[ht]  
 \includegraphics[width=7.6 cm]{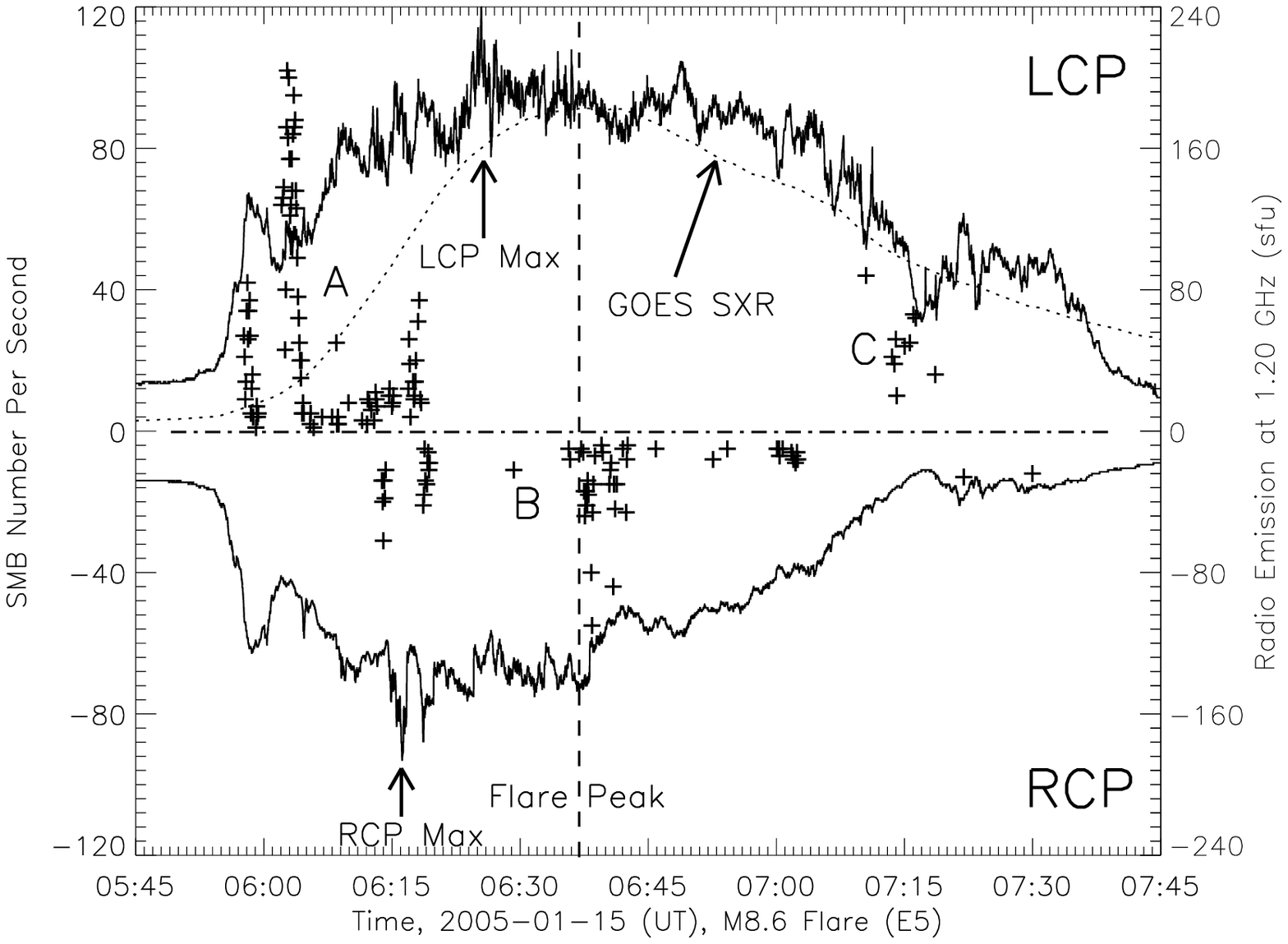}
 \includegraphics[width=7.6 cm]{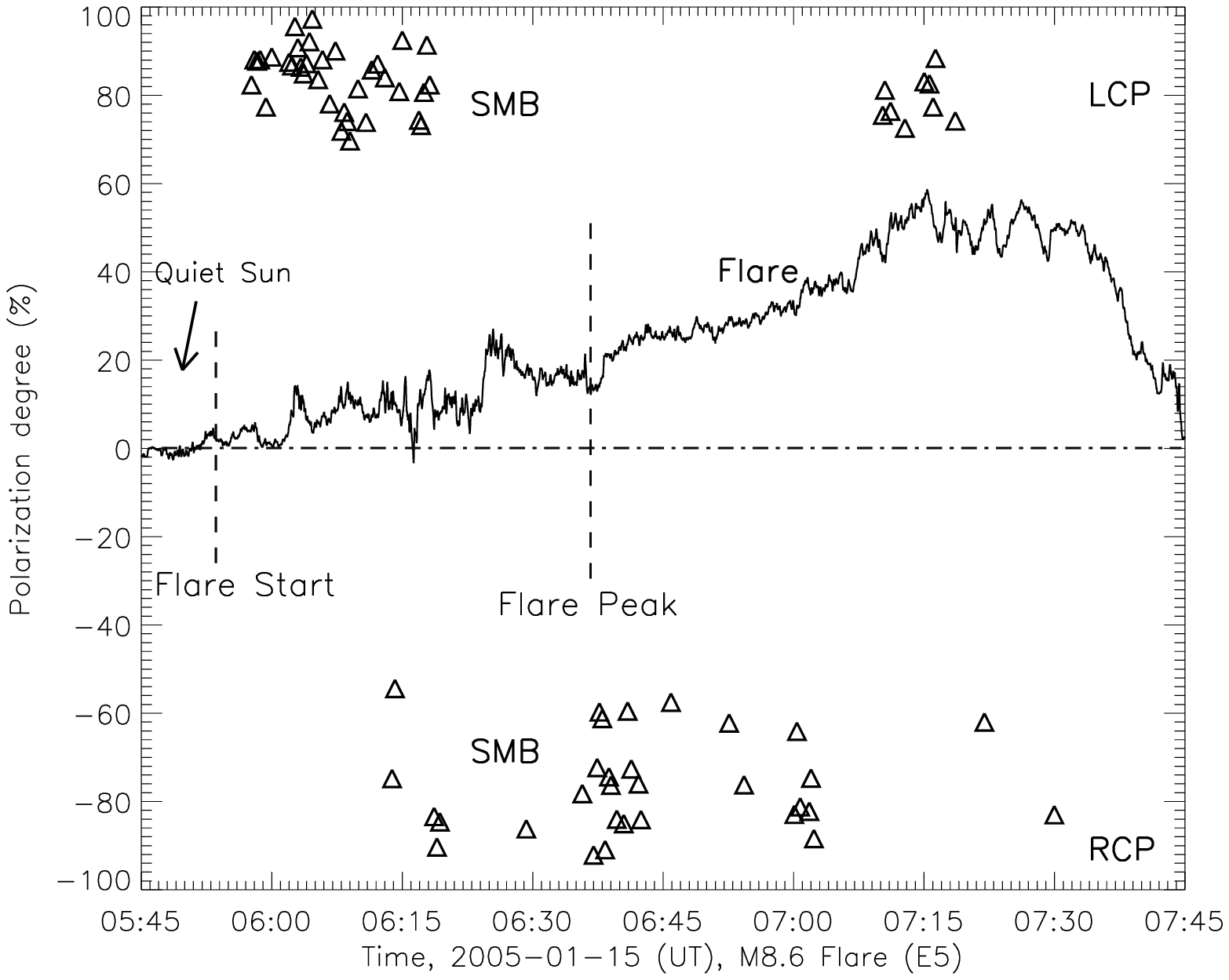}
\caption{The left panel is the distribution of SMB occurrence rate
in the M8.6 flare (E5) on 2005-01-15. The plus signs (+) indicate
the SMB occurrence rate. The solid curves are the radio emission
intensity at 1.20 GHz. The positive and negative values indicate
the presence of LCP and RCP, respectively. The dashed line is the
soft X-ray emission at 1-8 \AA~ observed by GOES. The right panel
is the comparison of the polarization degrees between the quiet
Sun (during 05:45-05:57 UT), flare (during 05:57-07:15 UT), and
SMBs ($\triangle$).} \label{fig:source}
\end{figure*}

The polarization is another important parameter to reflect the
nature of SMBs. The right panel of Fig.4 plots the temporal
distribution of the polarization degree of SMBs averaged in time
interval of one second (marked as $\triangle$) associated with the
M8.6 flare. In order to investigate the relationships between SMBs
and the underlying flare broadband emission, the polarization
degree of $F_{quiet}$ (during 05:45-05:57 UT) and $F_{flare}$
(during 05:57-07:15 UT) are also over-plotted on the same figure.
Here, we find that $F_{quiet}$ has no obvious polarization with
$r<10\%$, $F_{flare}$ has a moderate left-handed circular
polarization increasing slowly from the rising phase to decay
phase and in range of $r\simeq 10-55\%$, and then decreasing
rapidly to below 10\% after the end of the flare. Different from
the above two components, almost all SMBs are strong circular
polarization with degrees of polarization exceeding 55\%; it is
strong left-handed circular polarization before LCP maximum (06:25
UT) and strong right-handed circular polarization after LCP
maximum, mainly. There is no obvious correlation between the
polarizations of SMBs and background emission ($F_{quiet}$ and
$F_{flare}$). This fact implies that SMBs are independent bursts
overlapping on the underlying background emission.

\begin{deluxetable}{cccccccccccccc} 
\tablecolumns{9} \tabletypesize{\scriptsize} \tablewidth{0pc}
\tablecaption{List of solar flares observed by SBRS/Huairou in
NOAA 10720 during 2005 Jan 14-21\label{tbl-1}} \tablehead{
 \colhead{Event}& \colhead{Distribution}&\colhead{$F$(sfu)}&\colhead{$\tau$(ms)}&\colhead{$\triangle f$} &\colhead{$\triangle f/f$(\%)}&\colhead{$R_{drift}$}&\colhead{$N_{smb}$}&\colhead{L(km)}& \colhead{$T_{b}$(K)} &\\}
 \startdata
  E1   & rising phase &  44.3    & 71.6        &  35.5        &  3.0                &    820.5    & 90         & 600   & $1.28\times10^{12}$ \\\hline
  E5-A & rising phase &  52.8    & 15.1        &  23.6        &  2.0                &    3150     & 4600       & 400   & $3.46\times10^{12}$ \\
  E5-B & around peak  &  52.7    & 9.3         &  10.0        &  0.8                &    2120     & 1240       & 160   & $1.92\times10^{13}$ \\
  E5-C & decay phase  &  43.0    & 5.0         &  14.8        &  1.2                &    4110     & 150        & 240   & $7.16\times10^{12}$ \\\hline
  E10  & decay phase  &  39.0    & 221.1       &  41.7        &  3.5                &    152.9    & 45         & 700   & $8.18\times10^{11}$ \\
  E11  & rising phase &  40.1    & 49.7        &  24.8        &  2.1                &    736.3    & 92         & 420   & $2.38\times10^{12}$ \\
  E12  & rising phase &  46.0    & 69.5        &  32.6        &  2.7                &    965.0    & 17         & 540   & $1.58\times10^{12}$ \\\hline
  E13  & decay phase  &  55.9    & 10.3        &  20.0        &  1.7                &    3006     & 12600      & 340   & $5.10\times10^{12}$ \\
\enddata
\tablecomments{$\tau$ is lifetime, $\triangle f$ is frequency
bandwidth (MHz), $R_{drift}$ is the frequency drifting rate
(MHz/s). $N_{smb}$ is the SMB occurrence rate, L and $T_{b}$ are
the estimated upper limited width and lower limited brightness
temperature of the source region, respectively.}
\end{deluxetable}

According to SMB occurrence rates on the spectrograms, we may
partition them into three groups: group A occurred in the flare
early rising phase and mainly at LCP during 05:57-06:25 UT; group
B is occurred around the impulsive peak phase at RCP during
06:13-07:10 UT; and group C is occurred in the flare decaying
phase at LCP after 07:10 UT.

Among the 6000 SMBs, there are about 25\% SMBs which can be
measured obviously frequency drifting rates, the value of
$R_{drift}$ is in the range of 1200-9600 MHz/s. Table 2 lists the
averaged burst strength, lifetime, frequency bandwidth, and the
frequency drifting rate in each group of SMBs. Actually, the other
SMBs possibly also have frequency drifting rates which are not
detected because of the restrictions for the limited temporal and
spectral resolutions at the instruments, which implies that the
frequency drifting rates of SMBs may beyond 9600 MHz/s.

\begin{figure*}[ht]    
 \includegraphics[width=7.2 cm, height=8.0 cm]{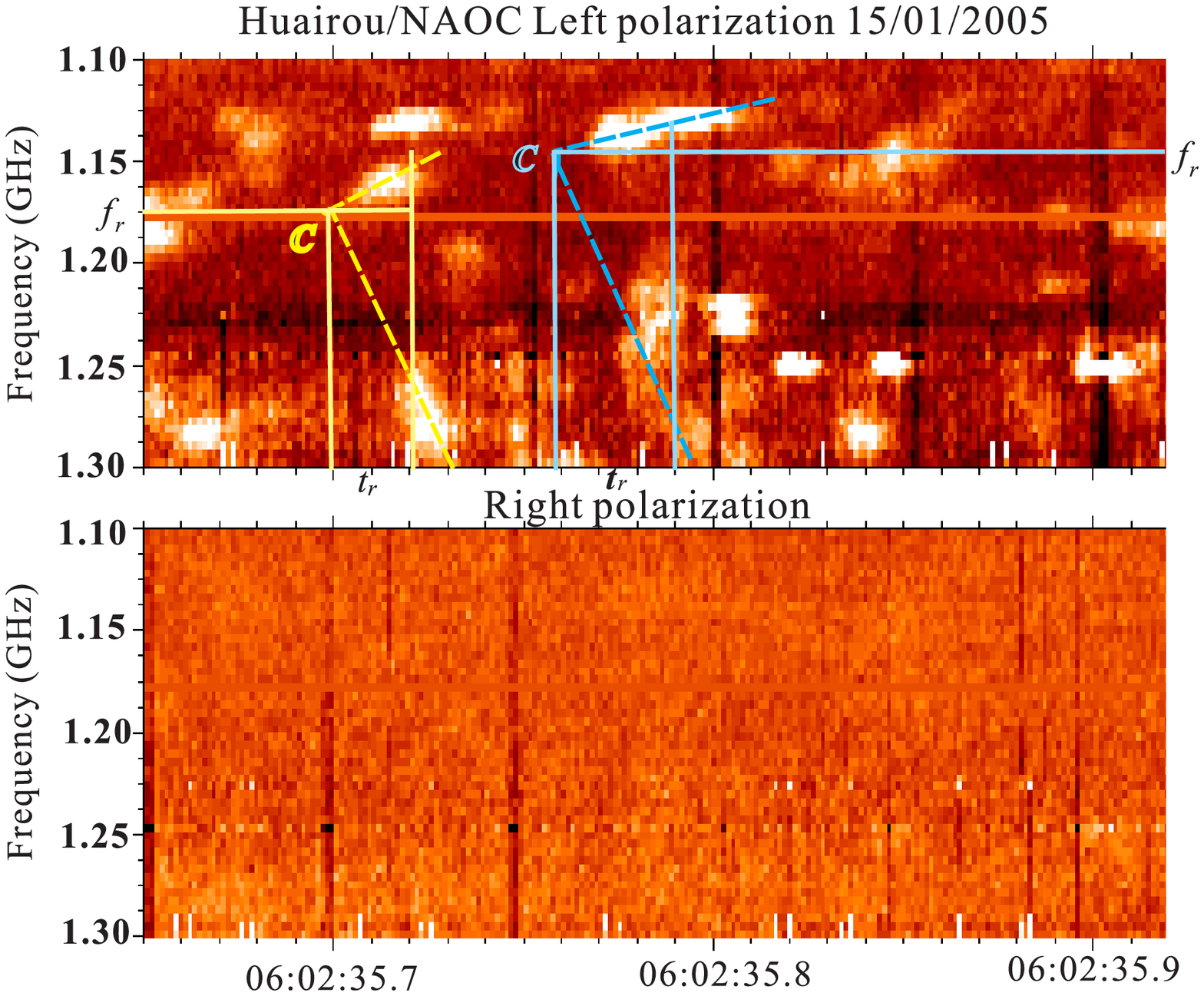}
 \includegraphics[width=7.2 cm, height=8.0 cm]{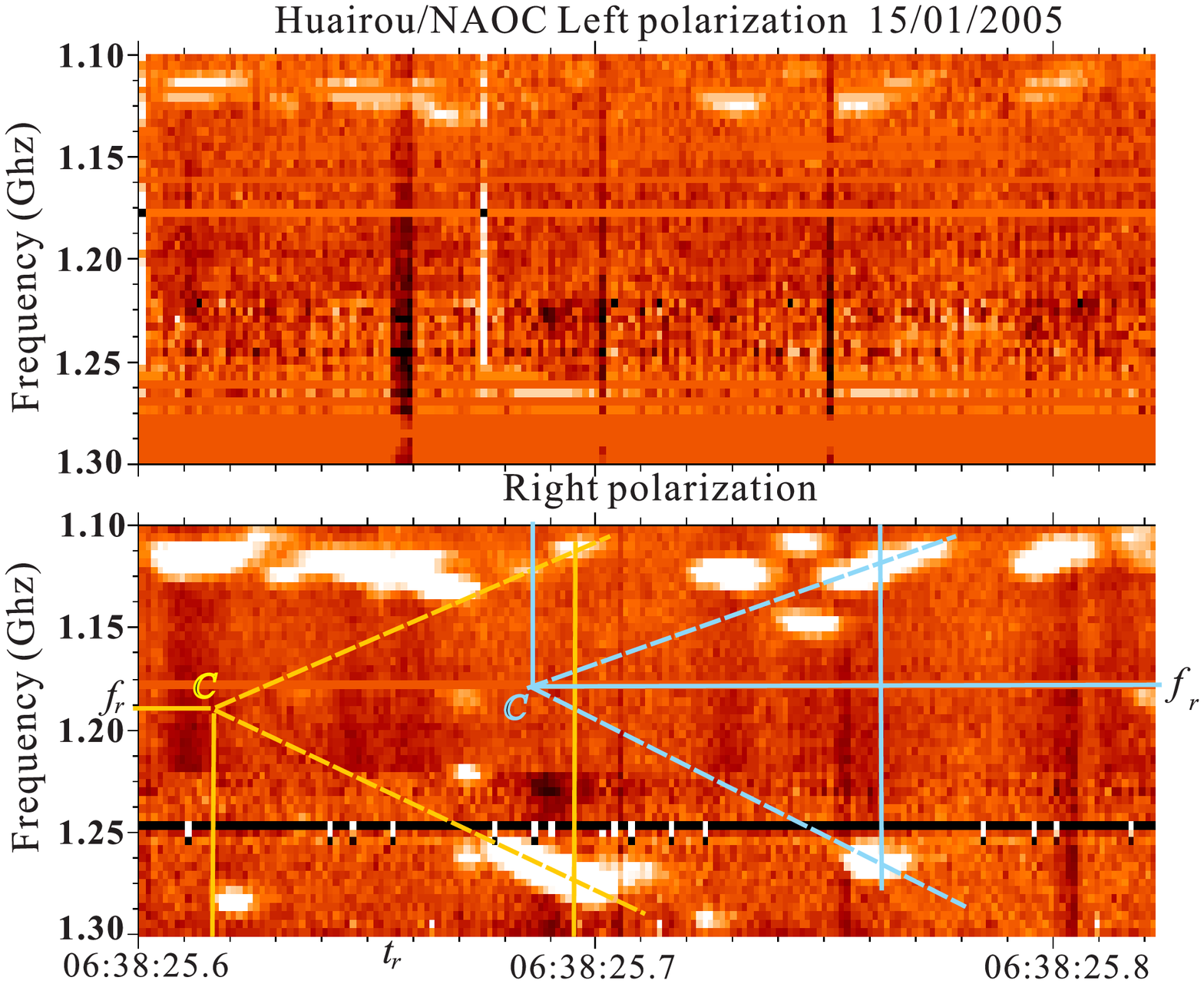}
\caption{Spectrograms of the reversed-drifting SMB pairs occurred
in the rising (left, 06:02:35.65-06:02:35.92 UT) and decay (right,
06:38:25.60-06:38:25.82 UT) phases of the M8.6 flare (E5) on
2005-01-15. The crossing point ($C$) of the dashed lines indicate
the reversed time ($t_{r}$) and a reversed frequency ($f_{r}$).}
\label{fig:source}
\end{figure*}

Sometimes, SMBs occurred in pairs with reversed frequency drifting
rates at nearly the same time and different frequency range, we
call them as reversed-drifting SMB pairs. Fig.5 presents two
examples of such reversed-drifting SMB pairs occurred at
06:02:35.65-06:02:35.92 UT (in the flare rising phase) and
06:38:25.60-06:38:25.82 UT (in the flare decay phase). The
crossing point of the extension lines (the dashed lines in Figure
5) of each SMB of the reversed-drifting SMB pairs can present a
reversed time ($t_{r}$) and a reversed frequency ($f_{r}$). Here,
$t_{r}$ is defined as the time difference between the crossing
point (C) and the corresponding SMB (approximately average of the
two SMBs), and $f_{r}$ is defined as the frequency where the
crossing point occurred. In the left panel of Figure 5, the
reversed-drifting SMB pair occurs at the left-handed circular
polarization, the reversed frequency $f_{r}\sim$1.150-1.175 GHz,
the low-frequency SMB is from 1.108 GHz to 1.168 GHz, with
frequency drifting rate of about -600 MHz s$^{-1}$; while the
high-frequency SMB is from 1.200 GHz to 1.300 GHz, with frequency
drifting rate of about 3900 MHz s$^{-1}$. The reversed time is
about 20-30 ms ahead of the corresponding SMBs. In the right panel
of Figure 5, the reversed-drifting SMB pair occurs at the
right-handed circular polarization, the reversed frequency
$f_{r}\sim$1.180-1.190 GHz, the low-frequency SMB is from 1.100
GHz to 1.150 GHz, with frequency drifting rate of about -800 MHz
s$^{-1}$; while the high-frequency SMB is from 1.250 GHz to 1.280
GHz, with frequency drifting rate of 1170 MHz s$^{-1}$. The
reversed time is 75-80 ms ahead of the corresponding SMBs. Among
all the reversed-drifting SMB pairs, the reversed frequency
$f_{r}\sim$1.15 - 1.19 GHz, and the negative frequency drifting
rate is in range from -500 MHz s$^{-1}$ to -3500 MHz s$^{-1}$,
while the positive frequency drifting rate is in the range from
1400 MHz s$^{-1}$ to 4000 MHz s$^{-1}$. Obviously, the
reversed-drifting SMB pairs in flare rising phase have a bit lower
reversed frequencies and much longer reversed time than that in
the flare decay phase. Supposing that SMB is associated with some
energetic electron beams in the ambient plasma, then the reversed
drifting SMBs may imply some local magnetic reconnections in small
scale regions, and the crossing points may indicate the electron
acceleration site where the electrons accelerated and propagated
upwards and downwards, and then triggered the formation of
reversed-drifting SMBs. Therefore, the reversed frequency may
reflect the region where magnetic reconnections take place, and
the reversed time may reflect the propagating process of energetic
electrons from the acceleration site to its emission source
region.

\subsubsection{Abundant SMBs in a long-duration X7.1 flare on 2005-01-20}

E13 is another SMB abundant long-duration flare. GOES soft X-ray
observation show that it starts at 06:36 UT, reaches to maximum at
07:01 UT, and ends at 07:26 UT on 2005-01-20 with duration 50 min.
It is an X7.1/2B flare, the most extremely powerful flare event
occurred in the deep descending phase of solar cycle 23. It has
attracted great attentions from solar and solar-terrestrial
community for its strong fast halo CME, strong gamma-ray bursts
with energy up to 200 MeV, the strongest solar energetic particle
(SEP) flux, and the strongest microwave bursts (Grechnev, et al,
2008; Bombardieri, et al, 2008; Wang, Zhao, \& Zhou, 2009). The
most important is that E13 is an SMB abundant event. From the
SBRS/Huairou observation at 1.10-1.34 GHz, there are about 12600
isolated SMBs distinguished. Different from E5, there is no any
isolated SMBs occurred in the early rising and impulsive peak
phase of the E13 flare. All SMBs are occurred after 07:10 UT, the
decay phase of the flare.

\begin{figure*}[ht]    
   \includegraphics[width=7.6 cm, height=8.8 cm]{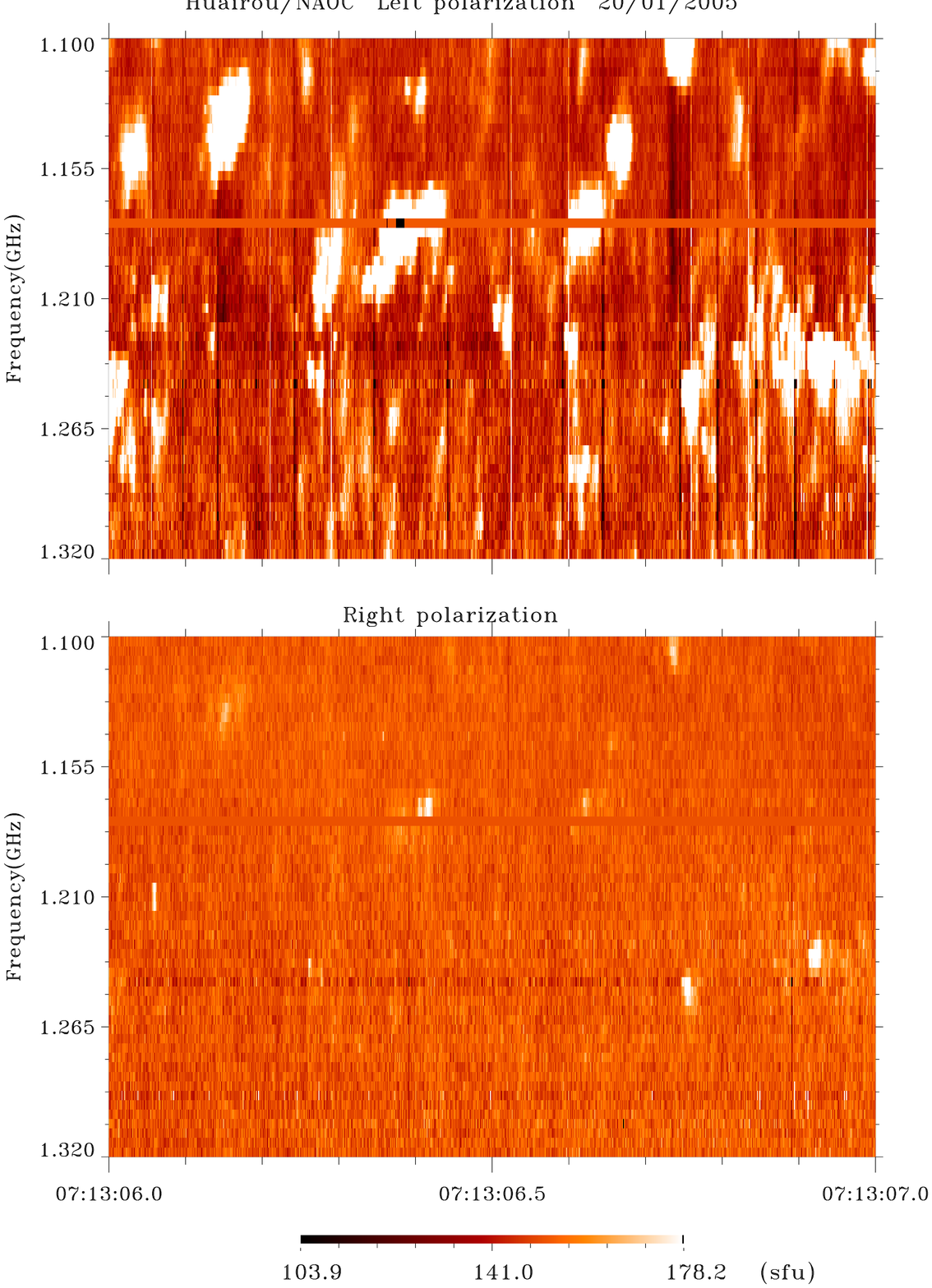}
   \includegraphics[width=7.6 cm, height=8.8 cm]{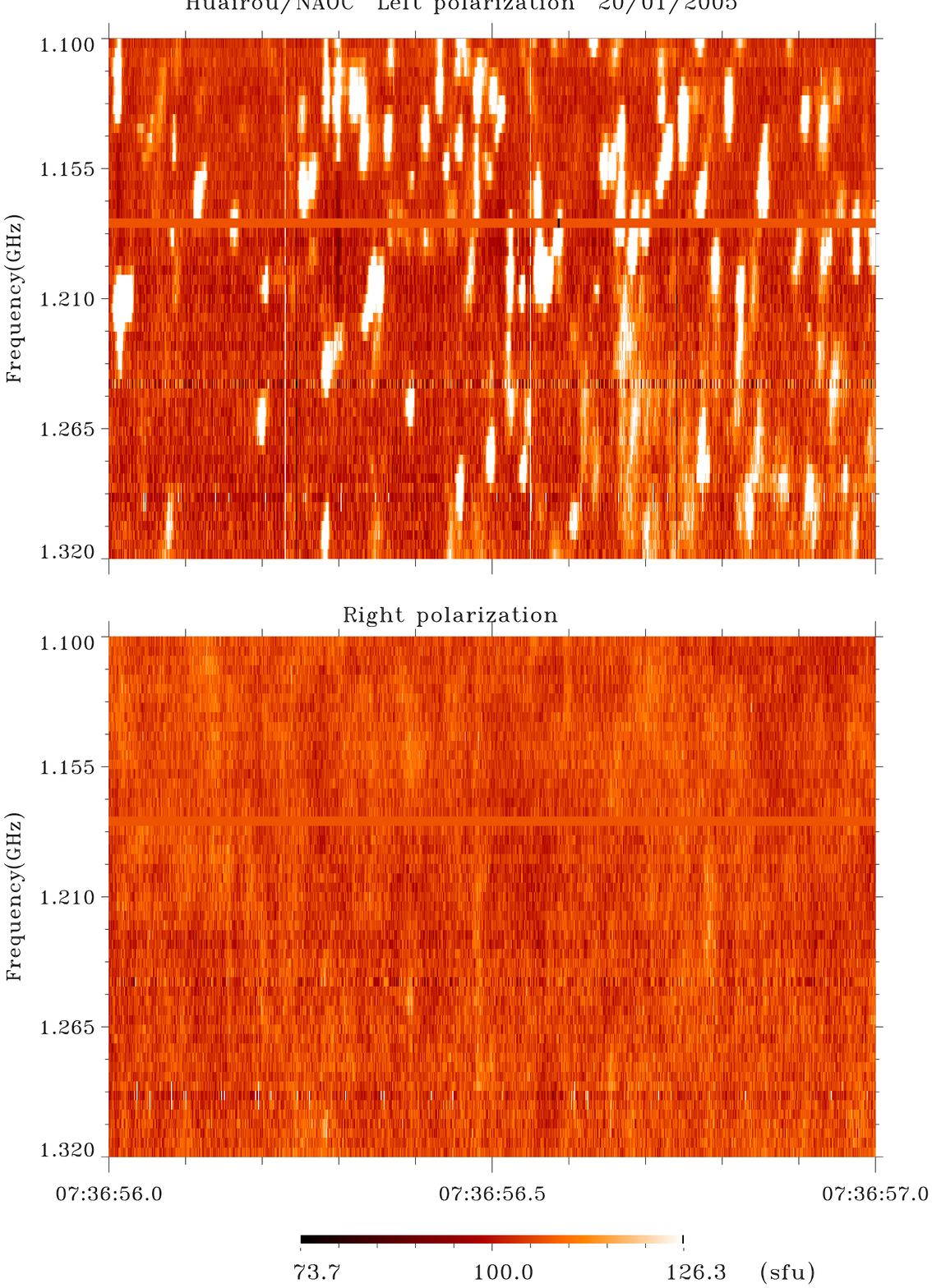}
\caption{Two segment spectrograms of SMBs at LCP and RCP observed
by SBRS/Huairou in the decay phase and after the ending of the
X7.1 flare (E13) on 2005-01-20.} \label{fig:source}
\end{figure*}

Fig.6 presents two segment spectrograms of SMBs at LCP and RCP
observed by SBRS/Huairou in the decay phase and after the ending
of the X7.1 flare (E13) on 2005-01-20. The left panels show a
one-second segment spectrograms of SMBs during 07:13:06-07:13:07
UT in the flare decay phase. There are about 30 SMBs with strong
left-handed circular polarization distributed randomly in this
segment. The frequency bandwidth of SMB is in the range of 12-40
MHz, and the lifetime is about 5-40 ms with averaged value 14 ms.
Some SMBs show frequency drifting, and $R_{drift}$ is in the range
from -2000 MHz/s to 3840 MHz/s. The underlying broadband flaring
continuum emission $F_{flare}$ is about 145 sfu with
$\sigma\simeq$ 6.4 sfu at LCP and 151 sfu with $\sigma\simeq$ 6.1
sfu at RCP. The burst strength of SMBs $F_{smb}$ is in the range
of 38-75 sfu with averaged value 55.8 sfu. The averaged
polarization degree is about 94\%. The right panels of Fig.6 show
another one-second segment spectrograms of SMBs during
07:36:56-07:36:57 UT after the flare ending. There are about 60
SMBs with strong LCP distributed randomly in this one-second
segment. The frequency bandwidth of SMB is in the range of 16-32
MHz with averaged value 24 MHz, and the averaged lifetime is about
9 ms. Some SMBs show the frequency drifting rate $R_{drift}$ from
2100 MHz/s to -3600 MHz/s. In this case, the underlying broadband
flaring continuum emission $F_{flare}$ is about 103 sfu with
$\sigma\simeq$4.7 sfu at LCP and 101 sfu with $\sigma\simeq$4.1
sfu at RCP. The burst strength of SMBs $F_{smb}$ is in the range
of 41-105 sfu with averaged value 59.5 sfu. The averaged
polarization degree is about 89\%. Similar to the SMBs occurred in
E5, there are also some SMBs occurred in E13 having reversed
frequency drifting rates. The reversed frequency is about 1.180
GHz, and the reversed time is about 90-105 ms ahead of the
corresponding SMBs.

\begin{figure*}[ht]     
 \includegraphics[width=7.6 cm]{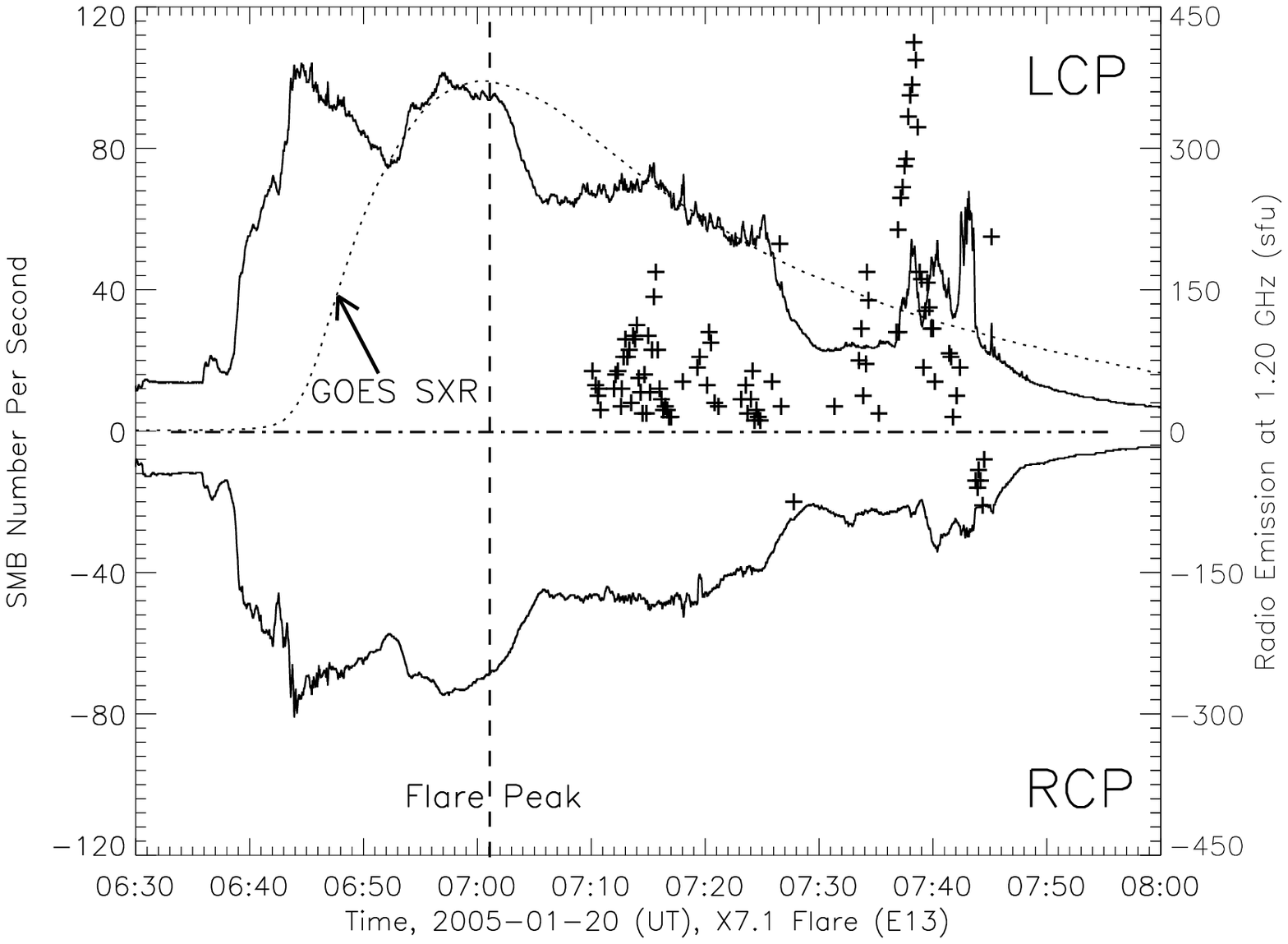}
 \includegraphics[width=7.6 cm]{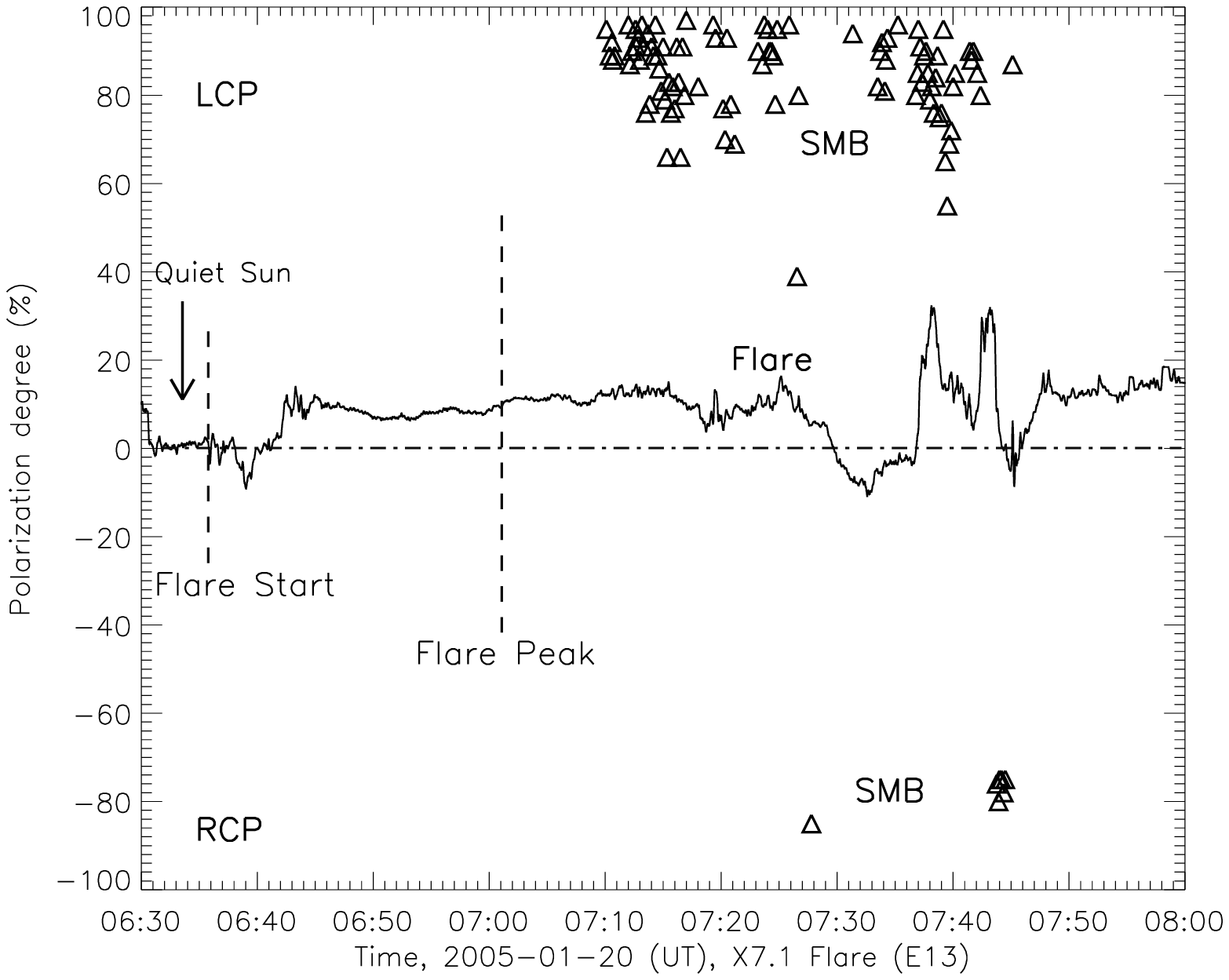}
\caption{The left panel is the distribution of SMB occurrence rate
in the X7.1 flare (E13) on 2005-01-20. The plus signs (+) indicate
the SMB occurrence rate. The solid curves are the radio emission
intensity at 1.20 GHz. The positive and negative values indicate
the presence of LCP and RCP, respectively. The dashed line is the
soft X-ray emission at 1-8 \AA~ observed by GOES. The right panel
is a comparison of polarization degrees between the quiet Sun
(before 06:36 UT), flare (during 06:36-07:26 UT), and SMBs
($\triangle$).} \label{fig:source}
\end{figure*}

From the spectrograms during the X7.1 flare, there are about 12600
isolated SMBs recognized at frequency of 1.10-1.34 GHz. The left
panel of Fig.7 presents the distribution of SMB occurrence rate in
the flare and comparisons with GOES solar X-ray emission, the
underlying microwave emission intensity at LCP and RCP. Here, we
find that all SMBs are occurred during 07:10-07:45 UT, after or
far away the maxima of GOES solar X-ray emission, underlying
microwave emission intensity at LCP and RCP. The SMB distribution
is very crowded together, and some of the SMB occurrence rate
($N_{smb}$) exceed 100, which is much higher than that occurred in
the above M8.6 flare event. The right panel of Fig.7 presents a
comparison of polarization degrees between the quiet Sun (before
06:36 UT), underlying flare broadband continuum emission (during
06:36-07:26 UT), and SMBs. Here, we can also find that the
polarization degree of the quiet Sun emission is very close to 0,
the underlying flare broadband continuum emission is weakly
left-handed circular polarization with $r\leq 20\%$, while most
SMB are strongly left left-handed circular polarization with
$r>60\%$. We can also find that a few SMB have strongly
right-handed circular polarization during 07:27 and 07:45 UT, far
way from the flare maximum. Here, once again, we find that SMBs
have obviously different polarization degree from the quiet Sun
and the underlying flaring broadband continuum microwave emission.
The range of SMB frequency bandwidth is 8-28 MHz, and the lifetime
is in 5-18 ms. There are also part of SMBs which can be detected
with obviously frequency drifting rates, and the value is in the
range of 1400-8800 MHz/s.

Table 2 lists the averaged burst strength, lifetime, frequency
bandwidth, and the frequency drifting rate of SMBs associated to
the X7.1 flare event.

\subsubsection{Sporadic SMBs in some flares}

Besides the above two SMB abundant flares, there are additional 4
flares having sporadic SMBs, including E1 (X1.2 flare, duration of
40 min), E10 (M6.7 flare, duration of 57 min), E11 (X1.3 flare,
duration of 37 min), and E12 (C4.8 flare, duration of 15 min) in
active region NOAA10720. Table 1 and Table 2 list the main
parameters of these events.

The left panels of Fig.8 present a 3.0 s segments of spectrogram
of a group of SMBs with LCP and RCP occurred in the early rising
phase of an X1.3 flare during 08:06:31 - 08:06:34 UT on 2005-01-19
(E11), which shows that SMBs are strongly right-handed circular
polarization with averaged $\bar{r}\simeq -77\%$. There are
totally 92 SMBs associated with the X1.3 flare, most of them have
frequency drifting rates in range from -205 MHz/s to -3200 MHz/s
with averaged value -736.3 MHz/s, no positive frequency drifting
rate. The burst strength is in the range of 29-72 sfu with
averaged value 40.1 sfu. The lifetime is in the range of 20-76 ms,
the averaged lifetime is 49.7 ms, and the averaged frequency
bandwidth is 24.8 MHz.

The right panels show a 2.3 s segment spectrogram of a few
isolated SMBs occurred during 07:43:19.7-07:43:22.0 UT in the
decay phase of another long-duration M6.7 flare (E10) on
2005-01-19. There are totally 45 isolated SMBs associated with the
flare. Some of them have frequency drifting rates in the range
from -61.3 MHz/s to -253.7 MHz/s. Here, SMBs are weakly right
polarization with $\bar{r}\simeq -31\%$, the averaged lifetime is
221.1 ms, the averaged burst strength $\bar{F}\simeq 39.0$ sfu,
and the averaged frequency bandwidth is 41.7 MHz which is much
wider than that occurred in the SMB abundant long-duration flares.

\begin{figure*}[ht]   
   \includegraphics[width=7.6 cm, height=8.8 cm]{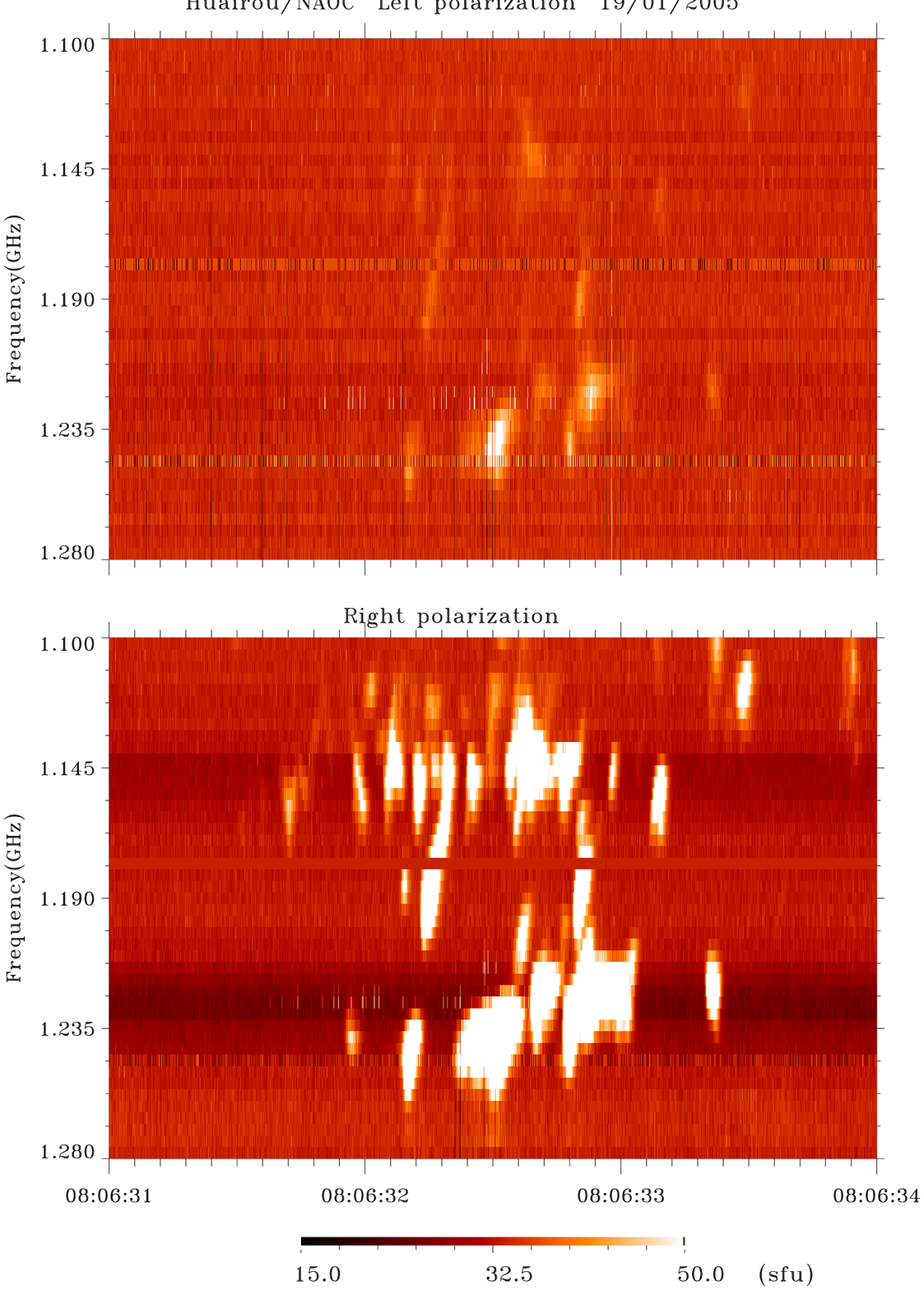}
   \includegraphics[width=7.6 cm, height=8.8 cm]{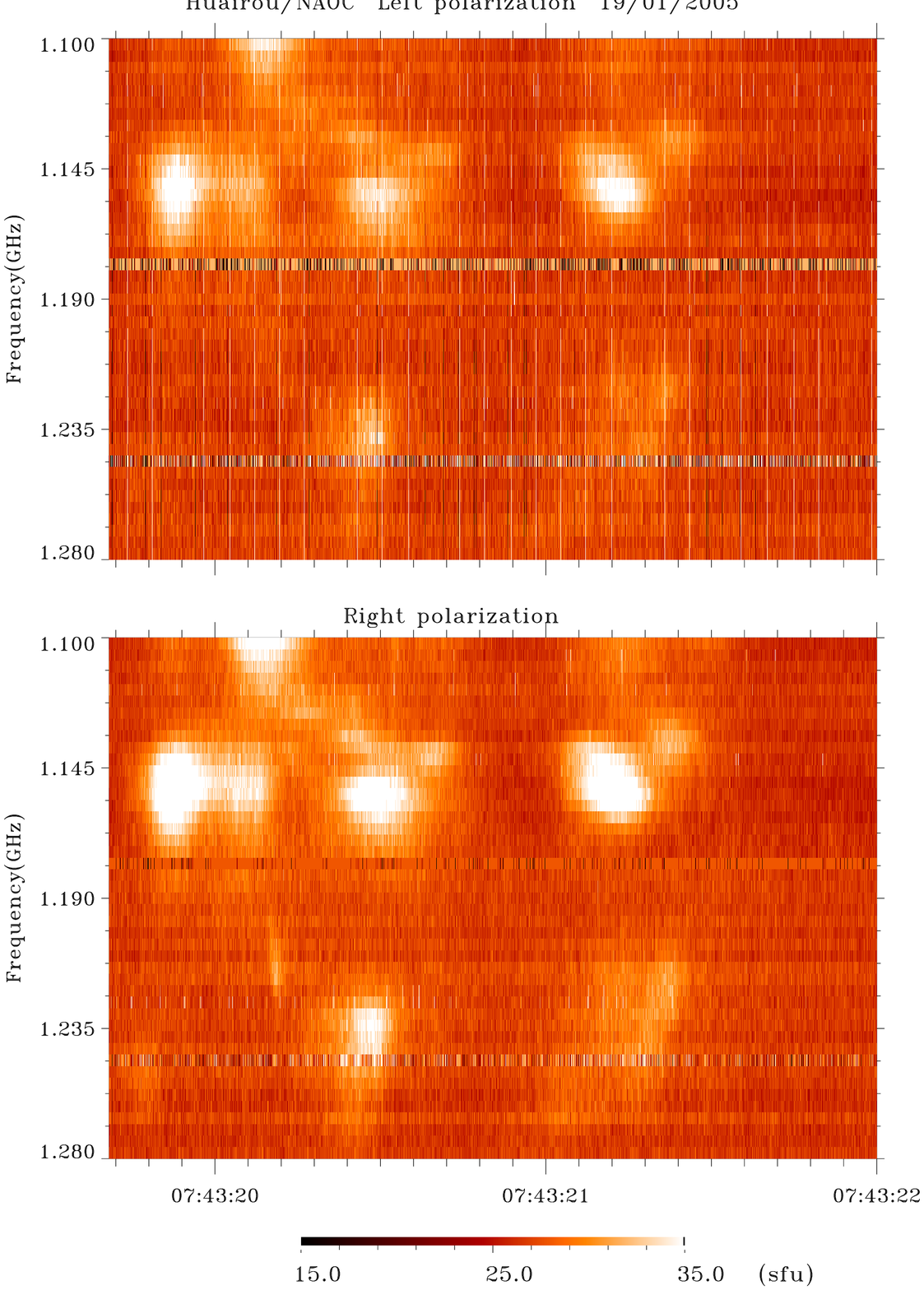}
\caption{The left panels present a segment of spectrograms of
sporadic SMBs in the rising phase of an X1.3 flare (E11) on
2005-01-19. The right panels present another segment spectograms
of sporadic SMBs in the decay phase of an M6.7 flare (E10) on the
same day.} \label{fig:source}
\end{figure*}

Actually, Table 2 indicates that in the 4 flare events with
sporadic SMBs, the SMB lifetime is much longer, the frequency
bandwidths are wider, the frequency drifting rates are much
slower, and the SMB burst strengths are a bit smaller than that in
the long-duration flares with a large amount of SMBs.

\subsubsection{Brief summary}

The above analysis shows that there are 6 flares in active region
NOAA10720 having isolated SMBs at frequency of 1.10-1.34 GHz, some
of them are abundant with SMBs, while others only have sporadic
SMBs.

(1) Among the 6 flare events, there are 3 events having SMBs in
their decay phase, and they are just the long-duration flares.
This fact implies that long-duration flares tend to produce more
SMBs in the decay phase.

(2) Most SMBs, especially the SMBs occurred abundantly in the
decay phase of long-duration flares, have strongly circular
polarization which is extremely different from the background
emission ($F_{quiet}$ and $F_{flare}$), very short timescale
($\sim$5-15ms), and narrow frequency band (around 0.8-2.0\%).

(3) In the long-duration flares, some SMBs are occurred in the
deep decay phase far away from the flare maxima, even after the
flare ending, and they still have similar burst strength and
strong polarization. Some of SMBs have reversed frequency drifting
rates, the reversed frequency is about 1.15-1.19 GHz, and the
reversed time is about 20-80 ms ahead of the corresponding SMBs.

(4) There is no obvious correlation of the burst strength,
lifetime, bandwidth, frequency drifting rates, polarizations
degrees between SMBs and background emission ($F_{quiet}$ and
$F_{flare}$). This fact implies that SMBs are independent bursts
overlapping on the underlying background emission.

\section{Physical Discussions}

From the relative frequency bandwidth of SMBs we may make an
estimation of the upper limited spatial scales of the source
region: $l\approx H_{f}\cdot\frac{\triangle f}{f}$. Here,
$H_{f}=|\frac{f}{\nabla f}|$ is the scale length of emission
frequency around the source region. The highly circular
polarization of SMBs indicate that the emission mechanism is
possibly the fundamental plasma emission, and the emission
frequency can be expressed: $f=9\sqrt{n_{e}}$. $n_{e}$ is the
electron plasma density. Then $H_{f}=2H_{n}$. Here,
$H_{n}=|\frac{n}{\nabla n}|$ is the scale length of plasma density
($n$) which depends on the plasma thermal temperature ($T_{e}$).
Supposing that thermal temperature around the source region is
about 1 MK, $H_{n}\sim 10^{4}$ km, and the source region is near
the lower part of corona. Then the upper limited spatial width of
SMB source regions can be obtained. The eighth column of Table 2
presents the estimated upper limited width of SMB source regions,
which is in the range of 160-700 km. This result is consistent
with the VLBI imaging observations which shows the diameter of
spike source region about 50 km at 1663 MHz (Tapping, et al,
1983). Furthermore, the lower limit of the SMB's emission
brightness temperature ($T_{b}$) can be deduced (Smerd, 1950; Tan
et al, 2009):
\begin{equation}
T_{b}\simeq 3.647\times10^{25}\frac{F_{smb}}{(\triangle f)^{2}}
(K).
\end{equation}
Here, units of $F_{smb}$ and $\triangle f$ are in sfu and Hz,
respectively. The last column of Table 2 lists the estimated lower
limited brightness temperature of SMBs, which is in range of
$8.18\times10^{11}$ - 1.92$\times10^{13}$ K. As a comparison, the
image observation indicates that the length of active region
AR10720 is about 200 arcseconds at 17 GHz (obtained from Nobeyama
Radio Heliograph), we may suppose that the source region of the
flaring radio emission will be larger than 200 arcseconds. As a
upper limit, the emission brightness temperature of the flare
radio emission at 1.20 GHz is about $4.5\times10^{7}$ K, which is
several orders lower than that of SMBs. The emission brightness
temperature of the quiet Sun can be also estimated as $6.3\times
10^{5}$ K, which is two orders lower than that of the flaring
broadband emissions.

The very high brightness temperatures of SMBs indicate that the
emission mechanism should be some coherent emissions, which may
reflect the presence of non-thermal energetic electrons in the
source region. Considering that they occur not only in the flare
rising and impulsive phases, but also in the flare decay phase and
even far away from the impulsive phase, possibly they reflect the
existence of similar small scale nonthermal energy releasing in
the flare decay phase.

The most frequently mentioned coherent mechanism is ECME which is
supposed to be the formation mechanism of solar radio spike bursts
by many people (Fleishman, et al, 2003, etc). However, ECME occurs
only when the following conditions are satisfied (Melrose \& Dulk,
1982):
\begin{equation}
\omega-\frac{s\omega_{ce}}{\gamma}-k_{\parallel}v_{\parallel}=0
\end{equation}
and
\begin{equation}
\omega_{ce}\gg\omega_{pe}
\end{equation}
Here, $\omega$ is the emission frequency, $s$ is the harmonic
number, $\gamma$ is the Lorentz factor of the energetic electrons,
$k_{\parallel}$ and $v_{\parallel}$ are the parallel components of
the wave number and electron velocity. The most favorable regime
to produce ECME is the loss-cone instabilities occurred near the
footpoint of flare loops. If the mechanism is ECME, then from
Equ.(3), the magnetic field strength in the SMB source region
should be $\gg$ 430 Gs.

So far, we have no direct measurement of magnetic field in solar
corona. Microwave zebra patterns (ZP) may be an useful tool to
provide information of magnetic field associated to the microwave
source regions (Tan et al. 2012). It is fortunate that there are
several ZPs accompanying with SMBs in E5 flare. Fig.9 presents
three segments of ZPs which occurred before the flare peak
(06:31:30-06:31:55 UT, left), in the flare decay phase
(06:51:06.4-06:61:07.6 UT, middle), and near the end of the flare
(07:17:30-07:17:34 UT, right), respectively. The left one is a
long-duration ZP, which frequency range is in 1.136-1.340 GHz
(central frequency is about 1.24 GHz), lasts for about 25 s, and
shows moderate RCP with polarization degree about 48\%. The
frequency separation between the adjacent zebra stripes is in the
range of 78-88 MHz, the average value is 84 MHz. The middle panel
shows an ZP at frequency range 1.10-1.25 GHz (central frequency
about 1.18 GHz), strong LCP with polarization degree of about
71.3\%, the frequency separation between the adjacent zebra
stripes is in the range of 16-20 MHz with averaged value 18 MHz.
The right panel shows an ZP at frequency range 1.10-1.20 GHz
(central frequency about 1.15 GHz), strong LCP with polarization
degree very close to 100\%. The frequency separation between the
adjacent zebra stripes is in the range of 12-16 MHz, the average
value is about 14 MHz. Fig.9 appears that the central frequency of
ZPs have an obvious decreasing from the early to late in the
flare. At the same time, the polarization of ZPs have a change
from RCP to LCP, which is approximately different from that of
SMBs. Using the similar method in Tan et al. (2012), the estimated
magnetic field strengths is about 126.8-143.0 Gs, 26-32.5 Gs, and
19.5-26.0 Gs, respectively in the source region, which is much
weaker than the requirement of Equ.(3).

\begin{figure*}[ht]   
 \includegraphics[width=5.4 cm]{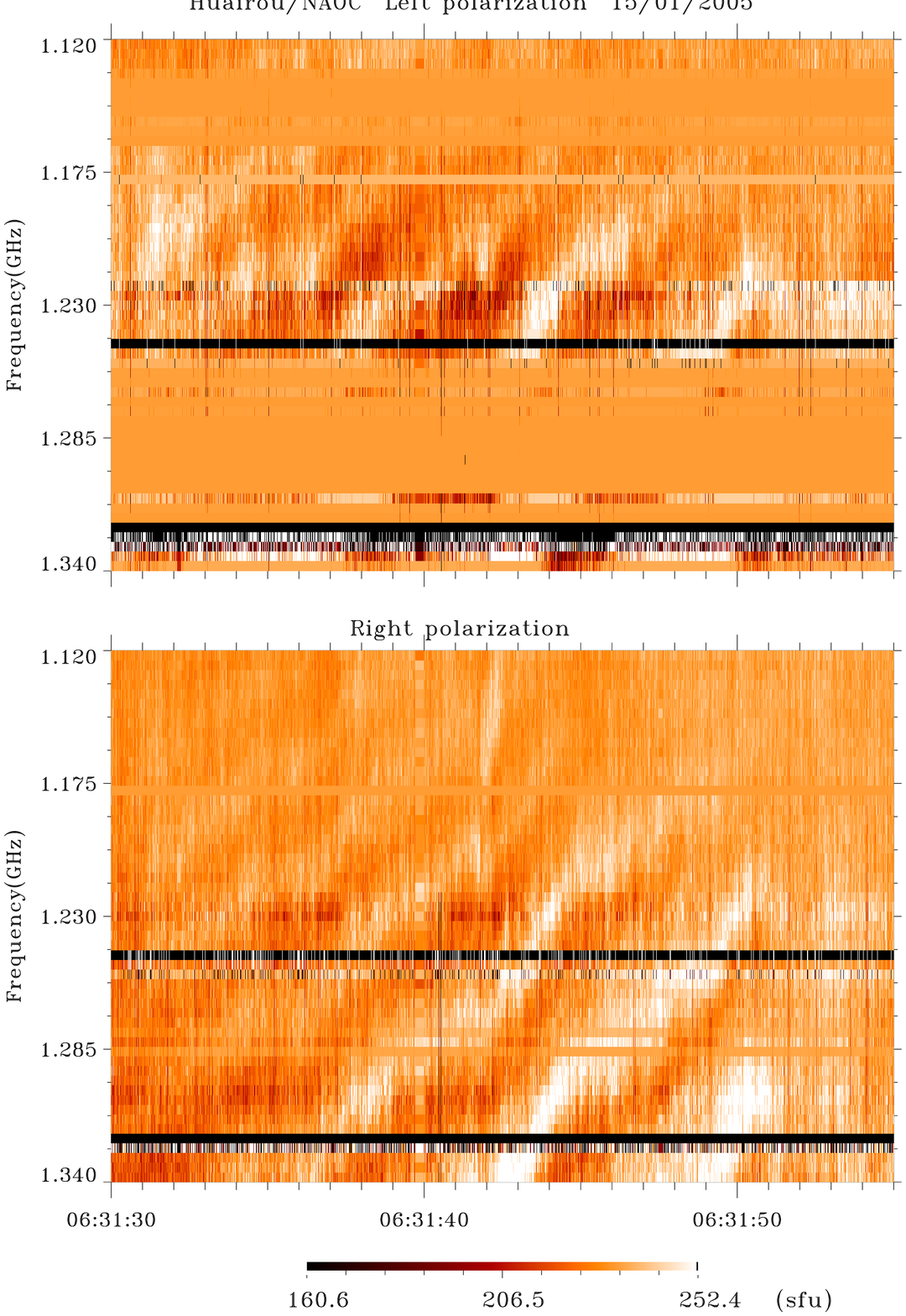}
 \includegraphics[width=5.4 cm]{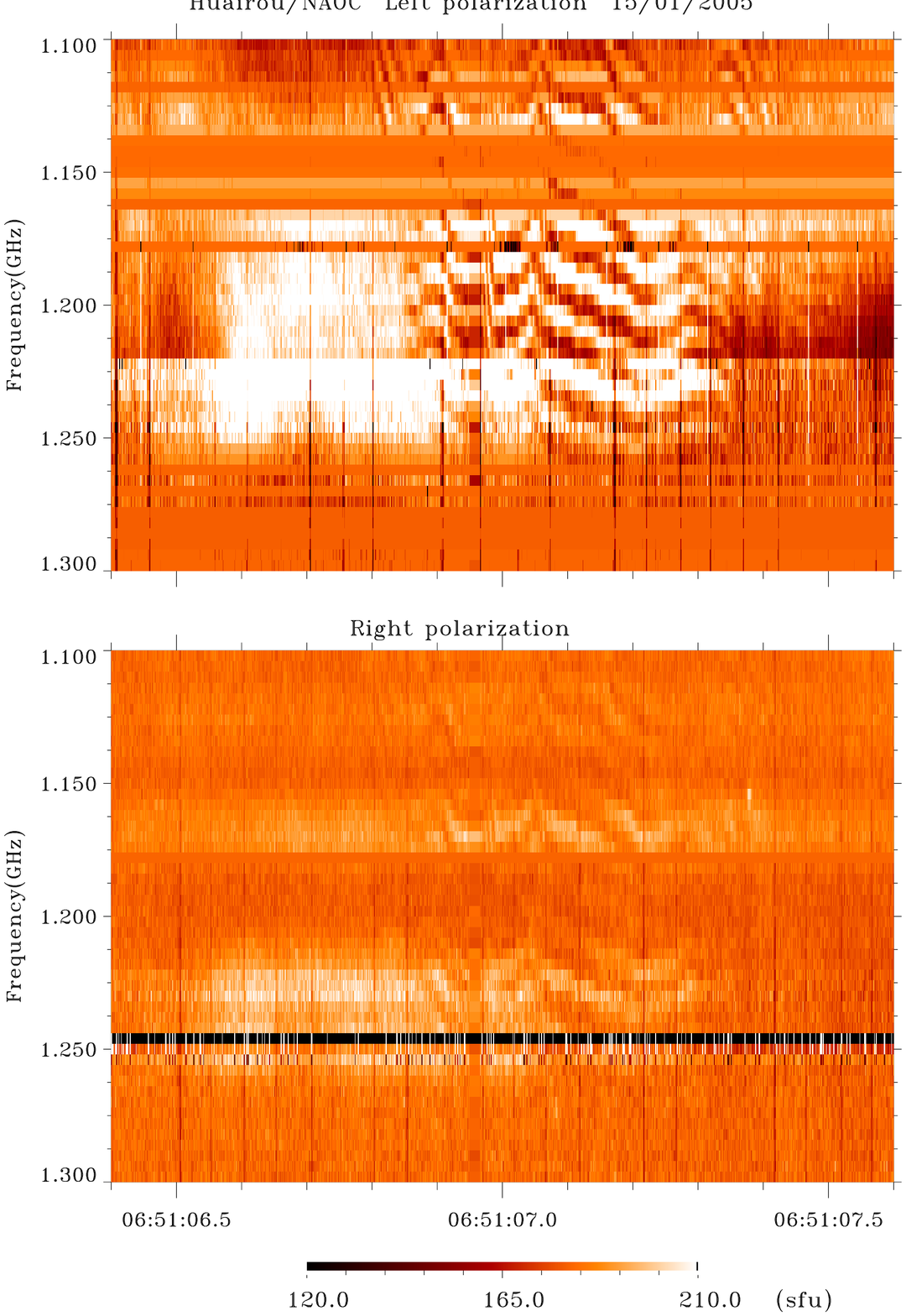}
 \includegraphics[width=5.4 cm]{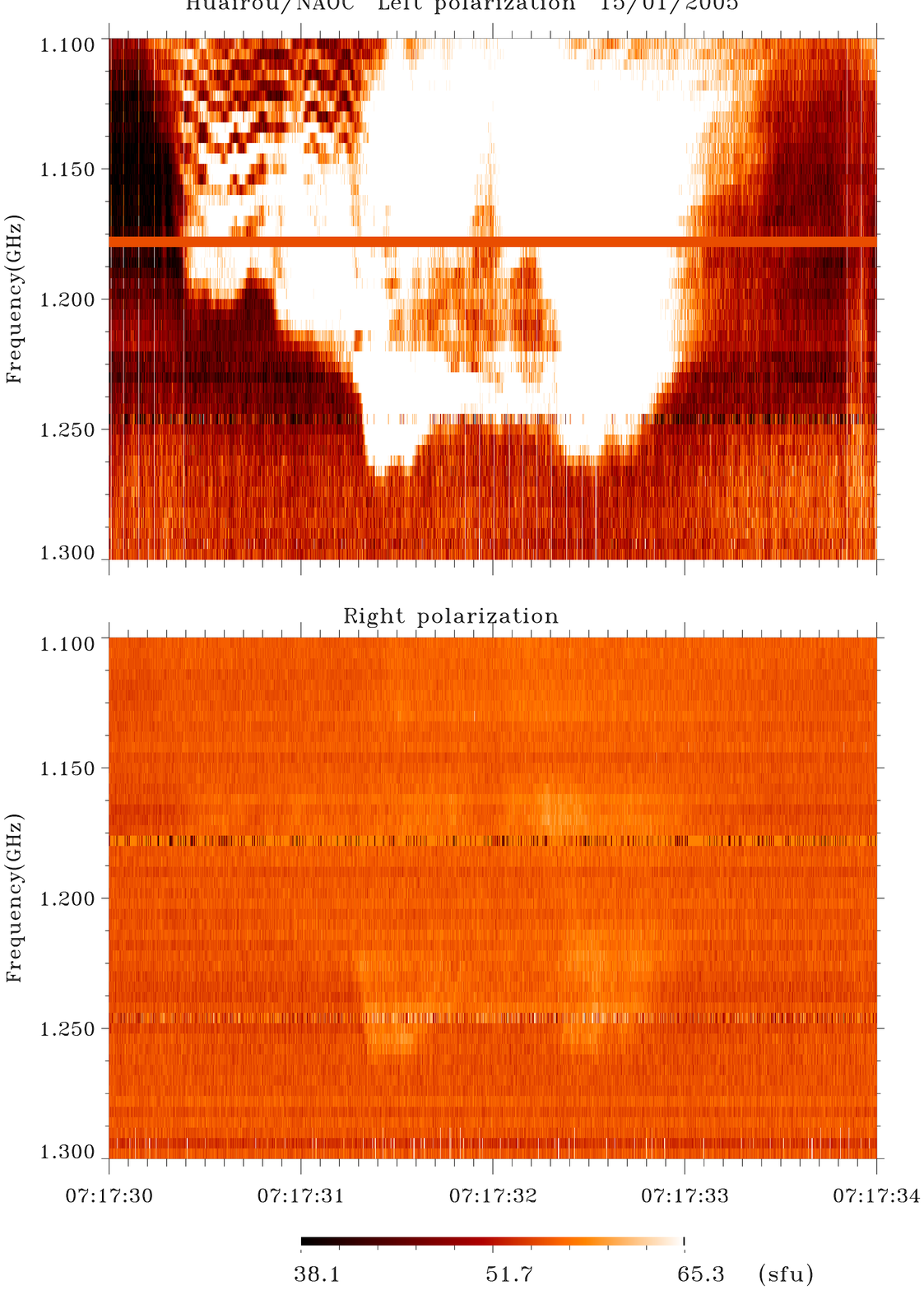}
\caption{The spectrograms of zebra patterns accompanying with the
M8.6 flare (E5) occurred in 06:31:30-06:31:55 UT (left),
06:51:06.4-06:61:07.6 UT (middle), and 07:17:30-07:17:34 UT
(right) on 2005-01-15, respectively.} \label{fig:source}
\end{figure*}

Actually, it is possible that the source region of SMBs may be
different from ZPs. However, the linear force-free magnetic
extrapolation indicates that it is very difficult to be more than
400 Gs in the lower corona region above the active region 10720
(Cheng, et al, 2011). Both results of the above estimation
indicate that ECME seems difficult to be the mechanism of SMB
formation.

The another kind of coherent mechanism is plasma emission, which
is generated from the coupling of two excited plasma waves at
frequency of $2\omega_{pe}$ with weak polarization, or the
coupling of an excited plasma wave and a low-frequency
electrostatic wave at frequency of about $\omega_{pe}$ with strong
polarization (Zheleznyakov \& Zlotnik 1975, Chernov et al. 2003).
The plasma waves are excited by fast electrons in relatively weak
magnetic fields:
\begin{equation}
v_{e}\gg v_{0}
\end{equation}
and
\begin{equation}
\omega_{pe}\gg\omega_{ce}
\end{equation}
Here, $v_{e}$ is the velocity of fast electrons, $v_{0}$ is the
velocity of the thermal electrons in background. Equ.(4) indicates
that energetic electrons are necessary, and fast electron beam
plays a dominant role in plasma emission processes. It is Langmuir
wave that can resonantly interact with electrons of $v\leq v_{0}$
and remove free energy. The recently only accepted form which can
produce high levels of Langmuir waves is the bump-in-the-tail
instability, or say two-streaming instability. There is only one
widely accepted means of creating bump-in-the-tail distribution:
having fast electrons outpace slow ones so that, at some distance
from the acceleration region, fast electrons arrive first, trigger
a Langmuir wave, resonantly interact with the ambient electrons,
and transfer the free energy into emission. When the Langmuir wave
reaches to some place where Equ.(5) does not agree, it will
quench. The very short lifetime of SMB requires that the source
region must be highly inhomogeneous filled with great numbers of
small scale structures. Similar to the regime of microwave
quasi-periodic pulsation with millisecond bursts (Tan \& Tan,
2012), here we may also adopt the same model to interpret the
formation of SMBs. We may suppose the post-flaring loops are
current-carrying plasma loops, where the resistive tearing mode
instability can be triggered, and causes the formation of many
small scale magnetic islands along each rational surface in the
plasma loops. Each X-point between the adjacent magnetic islands
will be a small reconnection site and will produce secondary
acceleration on the ambient electrons, similar to the regime in
current sheet (Shen, Lin \& Murphy, 2011). These accelerated
electrons impact the adjacent plasmas around the small X-point,
trigger the Langmuir turbulence and plasma waves, and produce
microwave bursts by plasma emission mechanism. Such microwave
bursts are just SMBs. As there are many magnetic X-points in the
current-carrying plasma loops, each X-point will be a small
reconnection site, and the region around each X-point will be a
source of an SMB. As a result, SMBs can arise in huge clusters.
However, as we are lack of imaging observations, it is a big
challenge to set up a more self-consistent structure of the SMB
source region.

The above analysis indicates that non-thermal energy releasing and
highly inhomogeneity in source region are necessary for the
formation of SMBs in the frame of plasma emissions.

As the frequency of plasma emission depends on the plasma density
$f=sf_{pe}\simeq 9sn_{e}^{1/2}$, when the fast electrons move in
the inhomogeneous ambient plasma, the frequency will drift:
$\frac{df}{dt}=\frac{9s}{2n_{e}^{1/2}}\frac{dn_{e}}{dr}\frac{dr}{dt}=\frac{f}{2H_{n}}v_{e}$.
Here, $f_{pe}=\frac{\omega_{pe}}{2\pi}$. $s$ is the harmonics,
$s=1$ is the fundamental emission, $s=2$ is the second harmonic
emission. Then the velocity of fast electrons can be estimated:
\begin{equation}
v_{e}\simeq 2H_{n}\cdot\frac{1}{f}\frac{df}{dt}
\end{equation}

Then, from the frequency drifting rates of SMBs, we may estimate
the velocity of the fast electrons. For example in the
long-duration flare E5, the frequency drifting rates are in the
range of 1200-9600 MHz s$^{-1}$, and the corresponding velocities
of fast electrons are in range of 0.07-0.53c, and the kinetic
energy is $3-80$ keV. Here, c is speed of the light. In another
long-duration flare E13, the frequency drifting rates are in the
range of 1400-8800 MHz s$^{-1}$, the corresponding velocities of
fast electrons are in range of 0.08-0.49c, and the kinetic energy
is $3-75$ keV. When $\frac{df}{dt}>0$, the exciter moves from
tenuous plasma region to dense plasma region, such as the downward
motion in corona; when $\frac{df}{dt}<0$, the exciter moves from
dense plasma region to the tenuous plasma region, such as the
upward motion in corona. The reversed-drifting SMB pairs may track
the position of acceleration region. From the reversed time we may
estimate the distance between the acceleration region and the SMB
source region: $D\simeq v_{e}t_{r}$. Supposing
$v_{e}\sim0.07-0.53c$, and $t_{r}\sim 20 - 80$ ms, then $D$ is
about 400-38000 km. This agrees with the theoretical requirement
that Langmuir wave can be triggered in place at some distance from
the acceleration region. And these facts are consistent with a
similar recent result from EUV and hard X-ray observations (Liu,
Chen, \& Petrosian, 2013).

However, in flares with only sporadic SMBs (E1, E10, E11, and
E12), the frequency drifting rates of some SMBs are slow as 60 -
900 MHz s$^{-1}$, the above method presents the velocities are
only in the range of 0.003-0.05 c. Considering their relatively
longer lifetimes, wider frequency bandwidth, moderate circular
polarizations, and occurred mainly in the early rising phase of
mid- or short-duration flares, they are possibly associated with
some small scale plasma jets. However, this is just a guess for
lack of corresponding imaging observations.

\section{Conclusions}

In the previous work (Benz, 1986, etc.), since most of SMBs
(mainly indicated spike bursts) appear during the primary energy
release in solar flares, they are regarded as fragmented into
about ten thousands more or less single elementary flare bursts.
And at least part of the flaring energy is carried by fast
electrons released in these elementary flare bursts. However, this
work presents a comprehensive analysis on a series of solar flares
occurred in active region NOAA 10720 during 2005 Jan 15-20, and
indicates that long-duration flares can produce abundant
independent isolated SMBs in the flare decay phase as well as in
flare rising phase, some of them are occurred even in the deep
decay phase far away after the flare peak, and even after the
flare ending. These SMBs have strongly circular polarization which
is extremely different from the background emission ($F_{quiet}$
and $F_{flare}$), very short timescale (around 5-15 ms), and
narrow frequency band (around 0.8-2.0\%). Possibly, they are
independent bursts overlapping on the underlying background
broadband continuum emission. The inferred brightness temperature
is at least $8.18\times10^{11}$ - 1.92$\times10^{13}$ K, and the
obviously different polarizations of SMBs from the background
emission (the quiet Sun and flare emission) indicates that each
SMB should be individual independent strong coherent emission
burst which is related to some non-thermal energy releasing and
production of energetic particles. These facts imply the existence
of energetic particles and the strong nonthermal energy releasing
processes with small scales in the decay phase of long-duration
flares. It is meaningful to the prediction of space weather
events.

As for the formation of SMB, because the magnetic field strength
deduced from ZPs and nonlinear force-free field extrapolation
around the source region is too weak, ECME seems difficult to be
the formation mechanism of SMBs, while the plasma emission
mechanism may become the favorable candidate for the formation of
SMBs at fundamental Langmuir frequencies, although it also has
some unresolved problems, such as the pattern of highly
inhomogeneity in source regions. From the plasma mechanism, we may
deduce the velocities and kinetic energy of fast electrons
associated with SMBs. Using the observations of reversed drifting
SMBs we may track region of the electron acceleration and estimate
the distance between the acceleration region and the SMB source
regions, etc.

As energetic electrons coming from solar flares may create severe
impacts to the environment of solar-terrestrial space. The
activity of nonthermal processes and energetic electrons in the
decay phases of long-duration flares indicate that it is also
important to pay close attention to the impact of post-flare
activities of long-duration flares on the space weather events.
The study of SMBs may reveal some new principles of the energetic
non-thermal processes associated with solar flares, such as
particles accelerations, the detailed structure of source region,
and the mechanism of energy conversions, etc.

Because of the instrument limitations, this work just presents the
behaviors of SMBs in a relatively small frequency range (1.10-1.34
GHz). Fig.6 and other spectrograms show that it is most possible
to exceed the frequency range of 1.10-1.34 GHz. It is necessary to
extend the observational frequency to below 1.10 GHz and above
1.34 GHz greatly with superhigh cadence and high frequency
resolutions. Additionally, the imaging observations at the
corresponding frequencies are also most important for its ability
to provide directly the locations, geometrical structures, and
magnetic fields in the source region (Yan, et al, 2009). These
will help us to set up a much more self-consistent theoretical
model of solar SMBs.

\acknowledgments The author thanks the referee for helpful and
valuable comments on this paper. Thanks are also due to GOES,
NoRP, SGD, and SBRS/Huairou teams for the systematic data. This
work is mainly supported by NSFC Grant No. 11273030, 11221063, and
11211120147, MOST Grant No. 2011CB811401, and the National Major
Scientific Equipment R\&D Project ZDYZ2009-3. This research was
also supported by Marie Curie Actions IRSES-295272-RADIOSUN.


\begin{thebibliography}{}
\bibitem[Battaglia(2009)]{Battaglia2009} Battaglia, M., Benz, A.O., 2009, \emph{A$\&$A}, \textbf{499}, L33

\bibitem[Benz(1985)]{Benz1985}Benz, A.O.: 1985, \emph{Solar Phys.}, \textbf{96}, 357.

\bibitem[Benz(1986)]{Benz1986}Benz, A.O.: 1986, \emph{Solar Phys.}, \textbf{104}, 99.

\bibitem[Benz(1991)]{Benz1991}Benz, A.O., Gudel, M., \& Isliker, H., et al.: 1991, \emph{Solar Phys.}, \textbf{133}, 385.

\bibitem[Benz(2002)]{Benz2002}Benz, A.O., Saint-Hilaire, P., \& Vilmer, N: 2002, \emph{A$\&$A}, \textbf{383}, 678.

\bibitem[Bombardieri(2008)]{Bombardieri2008} Bombardieri, D.J., Duldig, M.L., Humble, J.E., \& Michael, K.J.: 2008, \emph{ApJ}, 682,
1315

\bibitem[Cheng(2010)]{Cheng2010} Cheng, X., Ding, M.D., \& Guo, Y., et al.: 2010, \emph{ApJ}, 716, L68

\bibitem[Cheng(2011)]{Cheng2011} Cheng, X., Zhang, J., Ding, M.D., Guo, Y., \& Su, J.T.: 2011, \emph{ApJ}, 732, 87

\bibitem[Chernov(2003)]{Chernov2003} Chernov, G. P., Yan, Y.H., \& Fu, Q.J., 2003, \emph{A$\&$A}, \textbf{406}, 1071

\bibitem[Csillaghy \& Benz(1993)]{Csillaghy1993}Csillaghy, A., \& Benz, A.O.: 1993, \emph{A$\&$A}, \textbf{274}, 487.

\bibitem[Dabrowski(2011)]{Dabrowski2011} Dabrowski, B.P., Rudawy, P., \& Karlicky, M.: 2011, \emph{Solar Phys.}, \textbf{273},
377

\bibitem[Dulk(1985)]{Dulk1985} Dulk, G. A.: 1985, \emph{Ann. Rev. Astron. Astrophys.}, \textbf{23}, 169

\bibitem[Fleishman(2003)]{Fleishman2003} Fleishman, G.D., Gary, D.E., \& Nita, G.M.,: 2003, \emph{ApJ}, 593, 571

\bibitem[Fu(1995)]{Fu1995} Fu, Q.J., Qin, Z.H., Ji, H.R., \& et al: 1995, \emph{Solar Phys.}, \textbf{160}, 97

\bibitem[Fu(2004)]{Fu2004} Fu, Q.J., Ji, H.R., Qin, Z.H. \& et al.: 2004, \emph{Solar Phys.}, \textbf{222}, 167

\bibitem[Grechnev(2008)]{Grechnev2008} Grechnev, V.V., Kurt, V.G., Chertok, I.M. \& et al.: 2008, \emph{Solar Phys.}, \textbf{252},
149

\bibitem[Gudel \& Benz(1990)]{Gudel1990}Gudel, M., \& Benz, A.O.: 1990, \emph{A$\&$A}, \textbf{231}, 202.

\bibitem[Huang(2005)]{Huang2005} Huang, G.L., \& Nakajima, H.: 2005, \emph{Astrophys. Space Sci}, \textbf{295}, 423

\bibitem[Huang(2012)]{Huang2012} Huang, J., \& Tan, B.L.: 2012, \emph{ApJ}, \textbf{745}, 186

\bibitem[Jiricka et al(1993)]{Jiricka93} Jiricka, K., Karlicky, M., Kepka, O., \& Tlamicha, A.: 1993, \emph{Solar Phys.} \textbf{147}, 203.

\bibitem[Karishan(2003)]{Karishan03}Karishan, V., Fernandes F.C.R., Cecatto J.R., \& Sawant, H.S., 2003, \emph{Solar Phys.}, \textbf{215}, 147

\bibitem[Liu(2013)]{Liu2013} Liu, W., Chen, Q.R., \& Petrosian, V.: 2013, \emph{ApJ.} \textbf{767}, 168L.

\bibitem[Malville (1967)]{Malville1967} Malville, J.M., Aller, H.D., \& Jensen, C.J.: 1967, \emph{ApJ.} \textbf{147}, 711.

\bibitem[Melrose(1982)]{Melrose1982} Melrose, D.B., \& Dulk, G.A.: 1982, \emph{ApJ}, \textbf{259}, 844

\bibitem[Messmer(2000)]{Messmer2000}Messmer, P., \& Benz, A.O.: 2000, \emph{A$\&$A}, \textbf{354}, 287.

\bibitem[Rozhansky(2008)]{Rozhansky2008}Rozhansky, I.V., Fleishman, G.D., \& Huang, G.L.: 2008, \emph{ApJ.} \textbf{681}, 1688.

\bibitem[Smerd(1950)]{Smerd1950} Smerd, S.F.: 1950, \emph{Aust. J. Sci. Res. A}, \textbf{3}, 34.

\bibitem[Sawant(2001)]{Sawant2001} Sawant, H.S., Subramanian, K.R., Faria, C., et al: 2001, \emph{Solar Phys.}, \textbf{200}, 167.

\bibitem[Shen(2011)]{Shen2011} Shen, C.C., Lin, J., \& Murphy, N.A.: 2011, \emph{ApJ}, \textbf{737}, 14

\bibitem[Tan(2012)]{Tan2012} Tan, B.L., \& Tan, C.M.: 2012, \emph{ApJ}, \textbf{749}, 28

\bibitem[Tan et al(2012)]{Tan et al 2012} Tan, B.L., Yan, Y.H., \& Tan, C.M., et al.: 2012, \emph{ApJ}, \textbf{744}, 166

\bibitem[Tan et al(2009)]{Tan et al 2009} Tan, B.L., Yan, Y.H., \& Tan, C.M., et al.: 2009, \emph{Sci. China Ser. A}, \textbf{52},
1765

\bibitem[Tpping(1983)]{Tapping1983} Tapping, K.F., Kuijpers, J., \& Kaastra, J.S., et al.: 1983, \emph{A$\&$A}, \textbf{122}, 177

\bibitem[Tarnstrom(1972)]{Tarnstrom1972} Tarnstrom, G.L., \& Philip, K.W.: 1972, \emph{A$\&$A}, \textbf{16}, 21

\bibitem[Wang(2009)]{Wang2009} Wang, J.X., Zhao, M., \& Zhou, G.P.: 2009, \emph{ApJ}, \textbf{690},
862

\bibitem[Wang(2008)]{Wang2008} Wang, S.J., Yan, Y.H., \& Liu, Y.Y., et al.: 2008, \emph{Solar Phys.} \textbf{253}, 133.

\bibitem[Yan et al(2002)]{Yan2002} Yan, Y.H., Tan, C.M., \& Xu, L., et al.: 2002, \emph{Sci. Chin. A Suppl.}, \textbf{45}, 89.

\bibitem[Yan et al(2009)]{Yan09}Yan, Y.H., Zhang, J., \& Wang, W., et al.: 2009, \emph{Earth. Moon. Planet} \textbf{104}, 97.

\bibitem[Zhao(2006)]{Zhao2006}Zhao, M., \& Wang, J.X.: 2006, \emph{IAU Symposium} \textbf{233}, 41.

\bibitem[Zheleznyakov(1975)]{Zheleznyakov75} Zheleznyakov, V.V., \& Zlotnik, E.YA.: 1975, \emph{Solar Phys.} \textbf{44}, 461.

\end{thebibliography}
\end{document}